# Beyond VQE and QPE: A Noise- and Sampling-Error-Tolerant Quantum Algorithm with Heisenberg-Limited Precision

Qing-Xing Xie[a], Zidong Lin[b], Yun-Long Liu[a], Yan Zhao[cd*]

[a] *Department of Physics, Hubei University, Wuhan 430062, P. R. China.*
[b] *Shenzhen SpinQ Technology Co., Ltd., 518043, Shenzhen, P. R. China*
[c] *College of Materials Science and Engineering, Sichuan University, Chengdu 610065, P. R. China.*
[d] *The Institute of Technological Sciences, Wuhan University, Wuhan 430072, P. R. China.*
[*] yanzhao@scu.edu.cn

## Abstract

This paper introduces Witnessed Quantum Time Evolution (WQTE), a novel quantum algorithm for efficiently computing the eigen-energy spectra of arbitrary quantum systems without requiring eigenstate preparation—a key limitation of conventional approaches. By leveraging a single ancillary qubit to control real-time evolution operators and employing Fourier analysis, WQTE enables parallel resolution of multiple eigen-energies. Theoretical analysis demonstrates that the algorithm achieves Heisenberg-limited precision and operates with only a non-zero wavefunction overlap between the reference state and target eigenstates, significantly reducing initialization complexity. Numerical simulations validate the algorithm's effectiveness in molecular systems (e.g., $H_4$ chains) and lattice models (e.g., Heisenberg spin systems), confirming that computational error scales inversely with maximum evolution time while maintaining robustness against sampling errors and quantum noise. Experimental implementation on an NMR quantum processor further verifies its feasibility in real-world noisy environments. Compared to existing quantum algorithms (e.g., VQE, QPE and their variants), WQTE exhibits superior circuit depth efficiency, resource economy, and noise resilience, making it a promising solution for eigen-energy computation on noisy intermediate-scale quantum (NISQ) devices.

# 1. Introduction

Quantum computing has emerged as a revolutionary computational paradigm that harnesses fundamental quantum mechanical principles to solve problems intractable for classical computers across multiple disciplines[1,2]. A particularly formidable challenge is the precise simulation of complex quantum systems, which remains computationally prohibitive for classical approaches. Quantum computers, however, exploit superposition and entanglement phenomena to natively represent quantum states, offering a fundamentally powerful framework for such simulations[3,4]. This potential has driven significant research efforts toward developing efficient quantum algorithms for eigenstate determination - a critical capability with broad applications in materials science, quantum chemistry, and condensed matter physics[5–7].

Several algorithmic approaches have been explored, each with distinct limitations: Quantum Phase Estimation (QPE)[8–10], while theoretically exact, demands deep circuits with numerous Hamiltonian evolution operators, posing substantial implementation challenges for current noisy hardware[11,12]. Adiabatic State Preparation (ASP) suffers from severe $1/E_{gap}^3$ scaling (where $E_{gap}$ is the system's smallest eigen-energy gap), leading to impractical deep circuits[13–15]. The Variational Quantum Eigensolver (VQE)[16–19] improves resource efficiency but is hindered by the dual challenges of designing effective ansatz and navigating complex non-convex optimization landscapes[20]. Nonunitary Time Evolution (NTE) methods, despite offering certain theoretical advantages, face fundamental implementation barriers since quantum gates (except measurements) are inherently unitary[21].

In this paper, we introduce WQTE (Witnessed Quantum Time Evolution), a novel non-variational quantum algorithm for efficient eigen-energy computation. Compared to existing methods[22–24], WQTE operates without eigenstate preparation—requiring only a reference state with non-zero overlap—thus bypassing variational optimization. It achieves Heisenberg-limited precision[25], uses only one ancillary qubit for shallower circuits, exhibits inherent resilience to sampling and circuit noise, and resolves multiple eigen-energies in parallel via suitable reference states. These advantages make WQTE

a robust and resource-efficient alternative on near-term quantum hardware.[26–28]

The structure of this work is organized as follows. Section 2 presents the theoretical foundations of the WQTE algorithm and provides a complete description of its quantum circuit implementation. In Section 3, we rigorously validate the method through extensive numerical simulations spanning both molecular systems and condensed matter physics models. Our comprehensive evaluation framework incorporates: (i) a systematic analysis of accuracy to determine fundamental precision boundaries, (ii) a detailed assessment of computational resource requirements, and (iii) robustness testing against practical error sources including compiler-induced artifacts, statistical sampling variations, and quantum circuit noise effects. Building upon these simulation-based insights, Section 4 reprots experimental results from executing the algorithm on a current noisy intermediate-scale quantum (NISQ) processors. These real-device tests specifically characterize the algorithm's tolerance to both intrinsic quantum noise and measurement-related errors. Finally, Section 5 concludes the paper with an outlook on near-term applications and proposes promising directions for future algorithmic enhancements.

## 2. Theory and Methodology

This section is structured into two main parts: Subsection 2.1 introduces the theoretical foundations of the WQTE algorithm, while Subsection 2.2 elaborates on its practical implementation framework.

### 2.1 Fundamental Theory

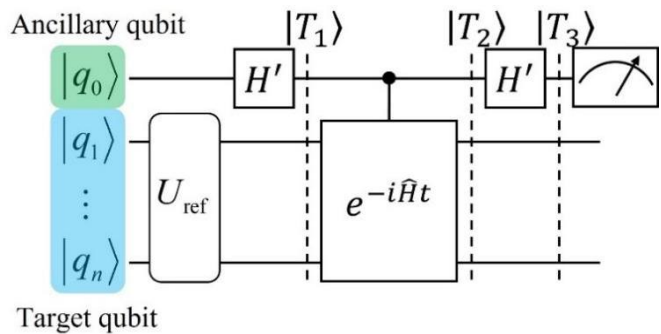

Figure 1. Quantum circuit for the WQTE algorithm. $q_0$ (highlighted within the green box) serves as the ancillary qubit, while qubits $q_1 \cdots q_n$ (enclosed in the cyan box)

are designated as target qubits.

The quantum circuit structure of WQTE is illustrated in Figure 1. An ancillary qubit, $q_0$, serves as the control qubit. The quantum circuit $U_{\text{ref}}$ is executed on target qubits $q_1 \cdots q_n$ to prepare the appropriate reference state wavefunction $|\psi_{\text{ref}}\rangle$. The choice of the reference state in the WQTE algorithm is flexible, as long as the reference state has a non-zero wavefunction overlap with the target eigenstate. For ground-state energy calculations in atomic and molecular systems, the Hartree-Fock(HF) state is often an excellent choice. When low-lying excited states are of interest, reference states can be selected from low-order excited configurations. In condensed matter models, such as the Heisenberg or Hubbard models, approximate solutions from classical computations can serve as effective reference states for the quantum circuit. Following this preparation, a Hadamard gate (labeled as $H'$ in the diagram, with the matrix form $H' = \frac{\sqrt{2}}{2}\begin{pmatrix} 1 & 1 \\ 1 & -1 \end{pmatrix}$) is operated on the ancillary qubit $q_0$. The Hadamard gate transform the ancillary qubit into the following state:

$$|T_1\rangle = \frac{\sqrt{2}}{2}|0\rangle \otimes |\psi_{\text{ref}}\rangle + \frac{\sqrt{2}}{2}|1\rangle \otimes |\psi_{\text{ref}}\rangle \tag{1a}$$

Subsequently, the ancillary qubit $q_0$ controls a real-time Hamiltonian evolution operator $e^{-i\hat{H}t}$, which acts on the target qubit register. The operator $e^{-i\hat{H}t}$ is applied only when $q_0$ is in the $|1\rangle$ state, resulting in the following outcome:

$$|T_2\rangle = \frac{\sqrt{2}}{2}|0\rangle \otimes |\psi_{\text{ref}}\rangle + \frac{\sqrt{2}}{2}|1\rangle \otimes e^{-i\hat{H}t}|\psi_{\text{ref}}\rangle \tag{1b}$$

Finally, a Hadamard gate is applied to the ancillary qubit $q_0$ once again, resulting in the output state $|T_3\rangle$:

$$|T_3\rangle = \frac{1}{2}|0\rangle \otimes (I + e^{-i\hat{H}t})|\psi_{\text{ref}}\rangle + \frac{1}{2}|1\rangle \otimes (I - e^{-i\hat{H}t})|\psi_{\text{ref}}\rangle \tag{1c}$$

By measuring Pauli-Z observable of the ancillary qubit in the output state $|T_3\rangle$, the probabilities of collapsing to states $|0\rangle$ and $|1\rangle$ are determined as follows:

$$\begin{aligned} P_0 &= \left\| \frac{1}{2}(I + e^{-i\hat{H}t})|\psi_{\text{ref}}\rangle \right\|^2 = \frac{1}{2} + \frac{1}{2}\langle\psi_{\text{ref}}|\cos(\hat{H}t)|\psi_{\text{ref}}\rangle \\ P_1 &= \left\| \frac{1}{2}(I - e^{-i\hat{H}t})|\psi_{\text{ref}}\rangle \right\|^2 = \frac{1}{2} - \frac{1}{2}\langle\psi_{\text{ref}}|\cos(\hat{H}t)|\psi_{\text{ref}}\rangle \end{aligned} \tag{2}$$

By varying the evolution time $t$ within the quantum processor, these collapse probabilities can be measured at various intervals, resulting in the function $Q(t) = P_0(t) - P_1(t)$, which simplifies to:

$$Q(t) = \langle \psi_{\text{ref}} | \cos(\hat{H}t) | \psi_{\text{ref}} \rangle \tag{3}$$

Assuming the Hamiltonian $\hat{H}$ possesses eigenstates $|\Psi_0\rangle, |\Psi_1\rangle, \cdots, |\Psi_i\rangle, \cdots$ with corresponding eigenvalues $E_0, E_1, \cdots, E_i, \cdots$, any reference state $|\psi_{\text{ref}}\rangle$ can be expressed as a linear combination of these eigenstates, with coefficients $\{c_i\}$:

$$|\psi_{\text{ref}}\rangle = \sum_i c_i |\Psi_i\rangle \tag{4}$$

Substituting into $Q(t)$ and noting that $\hat{H}|\Psi_i\rangle = E_i|\Psi_i\rangle$, we have:

$$Q(t) = \sum_i c_i c_j^* \langle \Psi_j | \cos(\hat{H}t) | \Psi_i \rangle = \sum_i |c_i|^2 \cos(E_i t) \tag{5}$$

Assuming each eigenvalue $E_i$ is an integer ($K_i$) multiple of a minimal unit $\delta$, such that $E_i = K_i \delta \quad K_i \in Z$, and $\delta$ is sufficiently small to make the quantization error negligible, it follows:

$$Q(t) = \sum_i |c_i|^2 \cos(K_i \delta t) \tag{6}$$

Additionally, evaluating $Q(t + 2\pi/\delta)$ results in:

$$Q\left(t + \frac{2\pi}{\delta}\right) = \sum_i |c_i|^2 \cos\left(K_i \delta t + K_i \delta \cdot \frac{2\pi}{\delta}\right) = Q(t) \tag{7}$$

This demonstrates that $Q(t)$ is a periodic function with period $T = 2\pi/\delta$. When expanded into a Fourier series, $Q(t)$ is expressed as:

$$Q(t) = \frac{a_0}{2} + \sum_{n=1}^{\infty} [a_n \cos(n\omega t) + b_n \sin(n\omega t)]$$

$$a_n = \frac{1}{T}\int_{-\frac{T}{2}}^{\frac{T}{2}} Q(t) \cos(n\omega t)\, dt \qquad b_n = \frac{1}{T}\int_{-\frac{T}{2}}^{\frac{T}{2}} Q(t) \sin(n\omega t)\, dt \tag{8}$$

where $\omega = 2\pi/T = \delta$, and given the even nature of $Q(t) = Q(-t)$, $b_n = 0$. It can be observed that Eq.6 and Eq.8 are formally identical. The Fourier coefficients $a_n$ directly correspond to $|c_i|^2$, and $n\omega$ to $K_i\delta$, essentially representing the eigenvalues $E_i$. Given the parity of the cosine function, $|n\omega| = |K_i\delta|$. Utilizing quantum computers, one can efficiently determine $Q(t)$ for any $t$. Subsequently, classical computers can be used to perform a Fourier transform to obtain the eigenvalues and calculate the overlaps between the eigenstates and the reference state.

Given that classical computers can only handle discrete signals and evaluating every time point within continuous intervals with quantum computers is impractical, it is necessary to define a reasonable sampling interval. This approach allows the measurement of $Q(t)$ at specific points, effectively converting a continuous signal into a format processable by classical computers. Sampling begins at $t = 0$, with an interval $\Delta$ and a total of $N$ samples: $t = 0, \Delta, 2\Delta \cdots N\Delta$. Quantum computers are utilized to capture $Q(t)$ values at each point, producing a series of discrete data points $q(n)$:

$$q(n) = Q(n\Delta) = \sum_i |c_i|^2 \cos(K_i \delta \cdot n\Delta) \tag{9}$$

The discretization of the time domain induces periodicity in the frequency domain, and vice versa. Let $R(k)$ represent the discrete Fourier transform (DFT) of $q(n)$:

$$R(k) = \sum_{n=0}^{N-1} q(n) e^{-i\frac{2\pi}{N}kn} \tag{10}$$

where $i$ represents the imaginary unit. The corresponding inverse Fourier transform is:

$$q(n) = \frac{1}{N} \sum_{k=0}^{N-1} R(k) e^{i\frac{2\pi}{N}nk} \tag{11}$$

The inverse transformation indicates that $q(n)$ is a periodic sequence with a period of $N$. Since $Q(t)$ satisfies $Q(t) = Q(-t)$, it follows that $q(n) = q(-n) = q(N-n)$. Substituting this into Eq. 10 demonstrates that $R(k)$ is an even function with a period of $N$. Based on the properties of the DFT, reconstructing the $q(n)$ sequence involves selecting $N$ consecutive $R(k)$ sequences starting from any point. For an odd number of samples, $N$, sample $R(k)$ from $-(N-1)/2$ to $(N-1)/2$:

$$q(n) = \frac{1}{N} \sum_{k=-\frac{N-1}{2}}^{\frac{N-1}{2}} R(k) e^{i\frac{2\pi}{N}nk} = a_0 + \sum_{k=1}^{\frac{N-1}{2}} a_k \cdot \cos\left(\frac{2\pi}{N}nk\right) \tag{12}$$

Where $a_0 = R(0)/N$ $a_k = 2R(k)/N$, comparing Eq. 9 and Eq. 12, and considering the parity of the cosine function, we find:

$$|c_i|^2 = a_i$$
$$\left|2\pi \cdot \frac{k}{N} n\right| = |K_i \delta \cdot n\Delta| \tag{13}$$

Eigenvalues are approximated as integer multiples of $\delta$, $E_i = K_i \delta$, with different $k$ values in $R(k)$ representing different multiples ($K_i$), thus indicating different eigenvalues. The corresponding equation is:

$$\frac{2\pi}{N} = \delta \cdot \Delta \qquad \frac{|k|}{N} = \frac{|E_i|}{2\pi} \cdot \Delta \qquad (14)$$

Therefore, based on the $\text{DFT}[q(n)]$ spectrum, if $a_i$ is non-zero, the corresponding frequency indicates an eigenstate with non-negligible overlap with the reference state, and its eigenvalues can be calculated using Eq. 14. It is worth noting that the quantum circuit sampling process introduces errors, and the quantization of eigenvalues $E_i$ is subject to inaccuracies. Therefore, when analyzing the frequency spectrum, emphasis should be on significant peaks, rather than all non-zero points.

## 2.2 Implementation Procedure

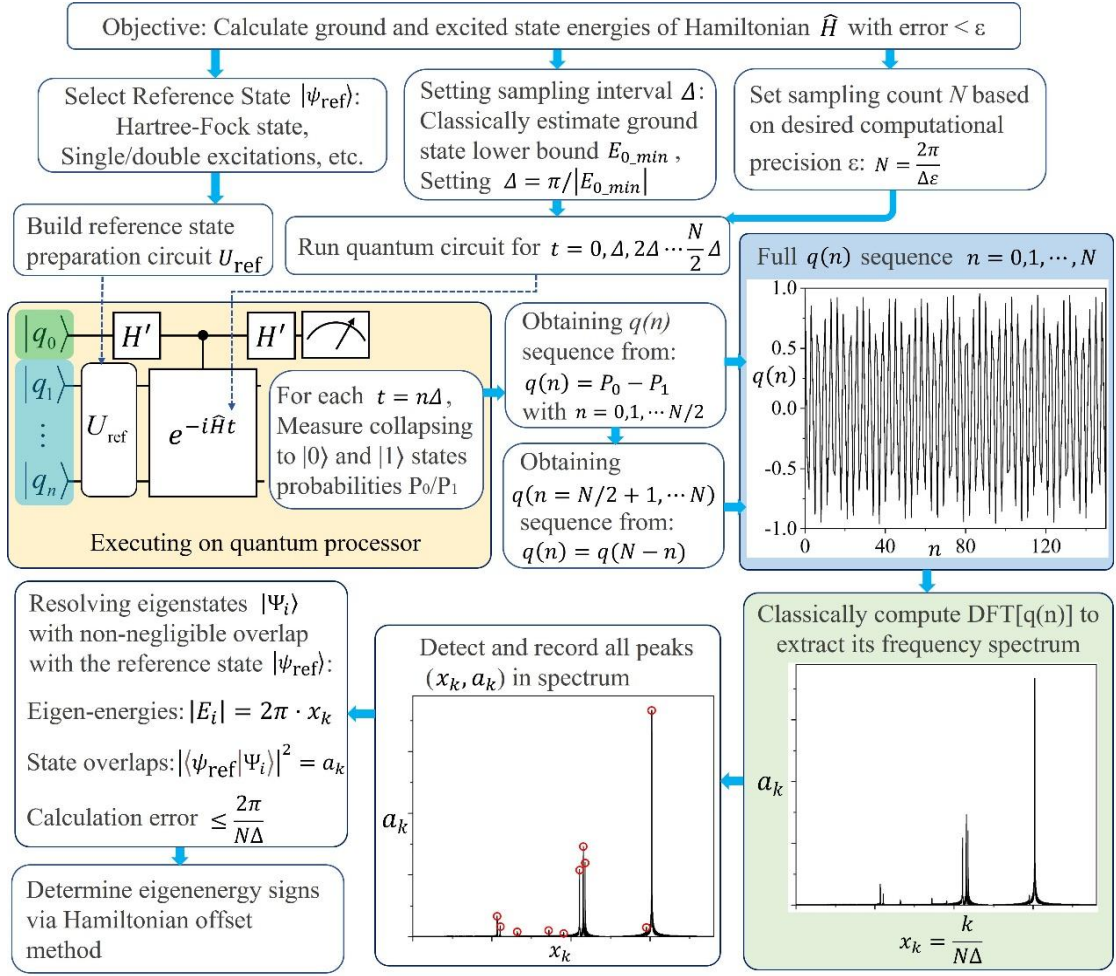

Figure 2. The computational workflow of the WQTE algorithm.

Figure 2 outlines the computational workflow of the WQTE algorithm for eigenvalue determination. The procedure begins with the input Hamiltonian $\hat{H}$ and target precision $\varepsilon$. An appropriate reference state is selected based on $\hat{H}$, requiring only non-vanishing overlap with target eigenstates, which allows for straightforward state preparation circuits ($U_{\text{ref}}$). The core principle for reference state selection is to maximize the wavefunction overlap with the target state (ground or excited state) based on research objectives, thereby minimizing the cost of quantum circuit sampling. Specific strategies have been supplemented in the manuscript as follows:

- Ground state calculation: For molecular systems, the Hartree-Fock (HF) state is preferred. For lattice models (e.g., Heisenberg or Hubbard models), classical approximate solutions (such as mean-field solutions or low-precision Density Matrix Renormalization Group solutions) are recommended.
- Low-lying excited state calculation: For molecular systems, Single/Double excitation configurations are suitable. Alternatively, trial wavefunctions for excited states can be constructed from ground-state reference states using quantum gates (e.g., X/Y gates), which is applicable to both molecular systems and lattice models.
- Variational quantum algorithm-assisted selection: Variational quantum algorithms can be used to obtain wavefunctions closely approximating the target state. For ground states, the VQE is adopted; for excited states, methods such as the Subspace-Search VQE[29], Variational Quantum Deflation[30], and Orthogonal State Reduction Variational Eigensolver[31] are feasible.

Concurrently, the sampling parameters are determined: under Nyquist sampling conditions (Sections 3.2 and 3.4), the precision $\delta = 2\pi/T_{\text{max}}$ (where $T_{\text{max}} = N\Delta$) depends solely on the maximum evolution duration $T_{\text{max}}$ rather than the sampling interval $\Delta$, which can be optimized as $\Delta = \pi/|E_{0\_min}|$ through classical estimation of the ground state energy lower bound $E_{0\_min}$ to minimize quantum resource requirements while maintaining ε-accuracy via $N = \frac{2\pi}{\Delta\varepsilon}$. The quantum implementation then proceeds by executing the circuit in Figure 1 for time steps $t = n\Delta$ ($n = 0,1,\cdots,N/2$), leveraging the even symmetry $q(n) = q(N - n)$ to reduce sampling costs, while measuring the q₀ qubit's |0⟩/|1⟩ collapse probabilities to construct the $q(n)$ sequence. Spectral analysis via discrete Fourier transform reveals eigenstates through significant peaks in $R(k)$, with corresponding eigenvalues calculated via Eq. 13, where peak amplitudes quantify state overlaps. Notably, quantum measurement noise necessitates focusing on dominant spectral features rather than all non-zero components. Finally, for eigenvalue sign determination (as detailed in Section 3.3), we implement a constant offset to the Hamiltonian and recalculate the eigenvalues. The signs of the original eigenvalues are then unambiguously determined by analyzing the systematic spectral shifts induced by this offset, where all eigenvalues exhibit consistent directional movement proportional to their signs.

## 3. Numerical simulation and discussion

In this section, we present numerical results to validate the proposed algorithm, beginning with an elaboration of key technical details and software implementations in Section 3.1. Section 3.2 not only verifies the algorithm's effectiveness but also systematically investigates the impact of time evolution sampling intervals ($\Delta$) through numerical simulations, offering theoretical guidance for practical parameter selection. Section 3.3 introduces a method for determining eigenvalue signs, while Section 3.4 examines the influence of maximum sampling duration ($T_{\max}$) on computational accuracy, establishing and numerically validating a quantitative model. Section 3.5 evaluates sampling errors and theoretically demonstrates the algorithm's inherent robustness: although errors may perturb spectral peak identification, they do not compromise the energy spectrum's intrinsic accuracy, with a systematic cost analysis provided for sampling overhead. Further analysis in Section 3.6 explores compilation errors and the relationship between circuit cost, system size, and computational precision. The impact of circuit noise is assessed in Section 3.7, where we prove the algorithm's resilience to noise to a certain extent. Building on these analyses, Section 3.8 consolidates scaling relations between computational costs (including qubit count, circuit complexity, and sampling requirements) and critical parameters such as system size and precision, with comparative benchmarking against leading quantum algorithms to highlight our method's advantages. Finally, Section 3.9 presents a comprehensive comparative analysis against state-of-the-art quantum algorithms.

## 3.1 Implementation Details of the WQTE Algorithm

To validate the effectiveness of the WQTE algorithm, we tested the method using the $H_4$ molecular system and the Heisenberg model. For the $H_4$ molecular chain, whose structure is depicted in Fig. 3(a), calculations were performed using the STO-3G basis set. The detailed procedure is as follows: First, molecular orbitals were calculated using the PySCF software[32–34], and the Hamiltonian was represented in its second quantized form. Then, we employ OpenFermion[35] to decompose both the reference state preparation unitary $U_{\text{ref}}$ and the time evolution operator $e^{-i\hat{H}t}$ into fundamental quantum gates, thereby constructing the explicit quantum circuit implementation depicted in Fig. 1. It is noteworthy that $\hat{H}$ is a multi-qubit Hermitian operator, which can often be expressed as a sum of Pauli string products: $\hat{H} = \hat{h}_1 + \hat{h}_2 + \cdots + \hat{h}_Z$. To compile the operator $e^{-i\hat{H}t}$, the Suzuki–Trotter[36–38] decomposition is generally employed. The first- and second-order Trotter decompositions are given by:

$$\hat{U}_1(t) = e^{-i\hat{h}_1 t}e^{-i\hat{h}_2 t}\cdots e^{-i\hat{h}_Z t} \tag{15a}$$

$$\hat{U}_2(t) = e^{-\frac{i\hat{h}_1 t}{2}}e^{-\frac{i\hat{h}_2 t}{2}}\cdots e^{-i\hat{h}_Z t}\cdots e^{-\frac{i\hat{h}_2 t}{2}}e^{-\frac{i\hat{h}_1 t}{2}} \tag{15b}$$

In contrast to the first-order Trotter expansion, the second-order expansion provides improved accuracy and additionally preserves time-reversal symmetry: $[\hat{U}_2(t)]^\dagger = \hat{U}_2(-t)$ and $\hat{U}_2(t)\hat{U}_2(-t) = \hat{I}$. Given that the collapse probability of the ancillary qubit derived in Eq. (2) leverages the time-reversal characteristic of the operator $e^{-i\hat{H}t}$, we employ the second-order Trotter expansion to approximate the operator $e^{-i\hat{H}t}$ during circuit compilation. Thirdly, an appropriate maximum sampling evolution time, $T_{\max}$, and sampling interval, $\Delta$, are set for the WQTE algorithm. According to the preceding analysis, it is theoretically assumed that all eigen-energies can be expressed as integer multiples of a minimum unit $\delta$, with the calculation accuracy determined by this quantization unit $\delta$. To ensure that the results fall within chemical accuracy, Eq. (14) implies that the maximum sampling evolution time $T_{\max}$ should satisfy:

$$\delta = \frac{2\pi}{T_{\max}} < 0.0016 \tag{16}$$

Hence, $T_{\max} > 3927$. For convenience, we set $T_{\max} = 4000$. By selecting different sampling intervals $\Delta$, the total evolution time sampling numbers $N$ can be determined by $N = T_{\max}/\Delta$. Fourthly, for each sampling point $t = n\Delta$, we substitute it into the quantum circuit shown in Fig. 1 and use MindSpore Quantum[39] platform to simulate. Finally, we perform spectral analysis of the simulated ancillary qubit collapse probabilities on a classical computer to identify their characteristic frequency components.

**3.2 Analysis of the Sampling Interval**

To demonstrate that this method can effectively calculate both the equilibrium and dissociation states of molecular systems, Figs. 3(a) and 3(b) display the computational results for the H$_4$ molecular chain with bond lengths of 1.0 Å and 2.0 Å, respectively, which correspond to the spectra of $q(n)$. Figs. 3(c) to 3(f) present the $q(n)$ spectra for various Heisenberg models, where the Hamiltonians of these systems can all be expressed as:

$$\hat{H} = J \sum_{\langle j,k \rangle} \left[ \sigma_j^x \cdot \sigma_k^x + \sigma_j^y \cdot \sigma_k^y \right] + h \sum_{\langle j,k \rangle} \sigma_j^z \cdot \sigma_k^z \qquad (17)$$

Here, $J$ represents the spin exchange coupling in the X and Y directions at each lattice site, $h$ denotes the coupling in the Z direction, and $\langle j, k \rangle$ refers to the nearest-neighbor lattice sites. The operators $\sigma_j^x$, $\sigma_j^y$, and $\sigma_j^z$ correspond to the three Pauli matrices at site $j$. Figs. 3(c) and 3(d) show the computational results for a one-dimensional Heisenberg model with 10 lattice sites, while Figs. 3(e) and 3(f) present the $q(n)$ spectra of a two-dimensional 4×3 Heisenberg model. The corresponding models and reference states are depicted in Figs. 3(c) and 3(e). For the Hamiltonians corresponding to Figs. 3(c) and 3(e), we have $J = h = 1$ (XXX model), while for Figs. 3(d) and 3(f), the parameters are $J = 1$  $h = 2$ (XXZ model).

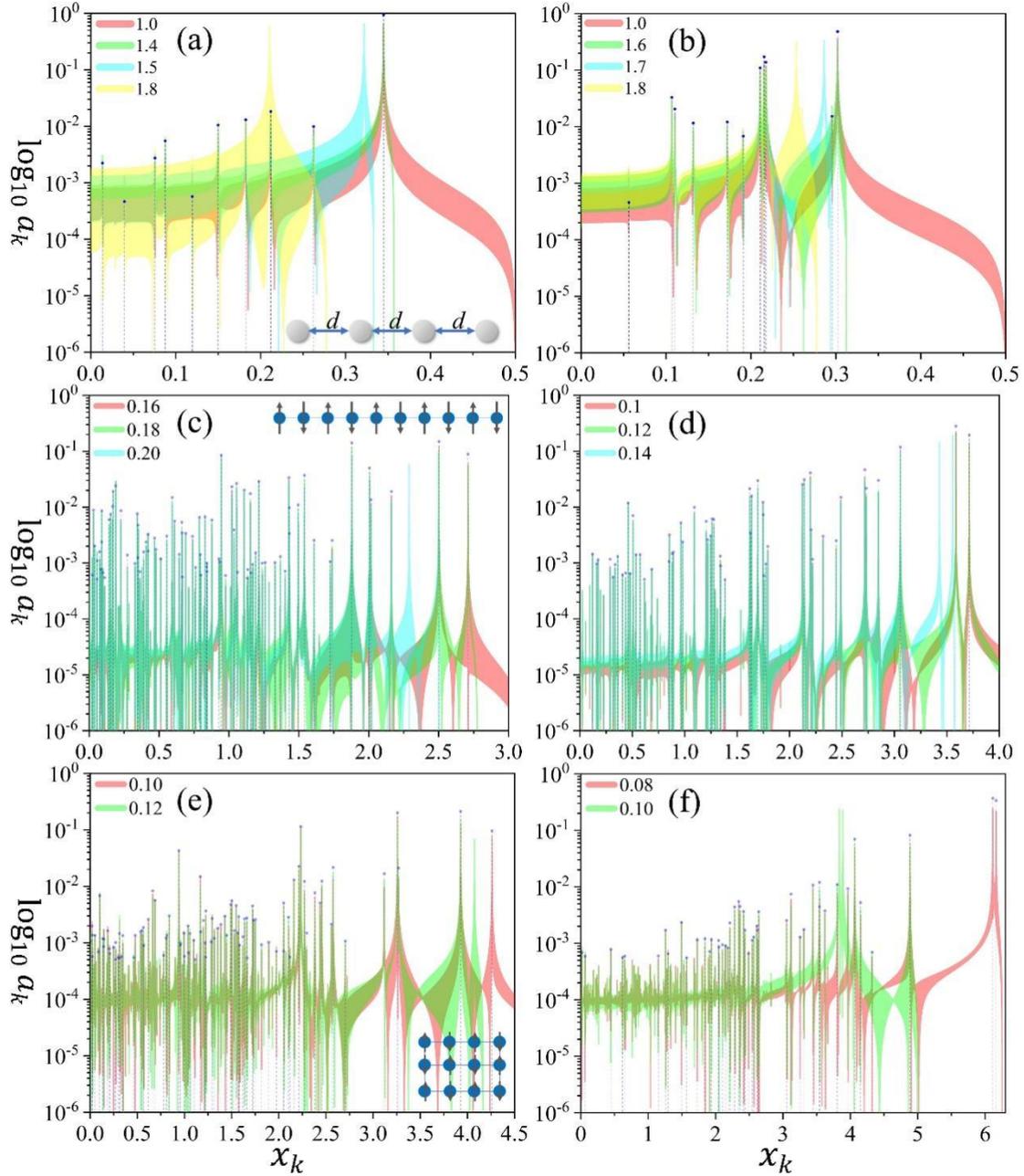

Figure 3. DFT spectra of the $q(n)$ signals for various quantum systems: H$_4$ molecular chain with bond lengths of 1.0Å(a) and 2.0Å(b), 10-sites of one-dimensional Heisenberg model with $J = h = 1$(c) and $J = 1\ \ h = 2$(d), 4×3 two-dimensional Heisenberg model with $J = h = 1$(e) and $J = 1\ \ h = 2$(f). The computational models are illustrated in (a), (c), and (e), with arrows at each lattice point in models (c) and (e) representing the Z-spin components, which are used to establish the reference state. Conversely, (a) and (b) employ the HF state as their reference state. The sampling time intervals ($\Delta$) are indicated in the legend. The x-axis is defined as $x_k = k/N\Delta$, and the y-axis represents $\log_{10} a_k$. The blue dotted vertical lines indicate the eigenstates obtained from the exact diagonalization method, where only eigenstates with overlaps greater than $5 \times 10^{-4}$ are displayed, i.e., eigenstates satisfying $|c_i|^2 \geq 5 \times 10^{-4}$ in Eq. 4. The x-coordinates of these lines are $|E_i|/2\pi$, and their heights are $|c_i|^2$.

The DFT spectra of the $q(n)$ signals of different quantum models with various sampling intervals are shown in Fig. 3. It is noteworthy that all $q(n)$ signals in Fig. 3 shares the same maximum sampling evolution time $T_{\max} = 4000$, hence different sampling intervals imply different total sampling numbers: $N = T_{\max}/\Delta$. As per the previous analysis, peak points in the spectra represent various eigenstates of the Hamiltonian. The x-coordinates denote different frequencies: $x_k = \frac{k}{N\Delta}$, while the y-coordinates represent the Fourier transform coefficients of different frequencies $a_k$. According to Eq. 14, $|E_i| = 2\pi x_k$. For the convenience of comparison, some eigenstates with non-negligible overlaps with the reference state are displayed using dotted vertical lines. These eigenstates are obtained via the exact diagonalization (ED) method. Theoretically, the positions of these lines will coincide with peaks in the spectrum.

Fig. 3 demonstrates that when the sampling interval is below the threshold, although the shapes of the spectrum curves may vary, the peaks of all curves align perfectly with the vertical lines representing the eigenvalue (ED) positions. However, when the sampling interval exceeds the threshold, the peaks no longer align with the ED lines. The threshold is system-dependent. For an H$_4$ molecular chain with a bond length of 1.0 Å, the sampling interval must not exceed 1.4. When the bond length increases to 2.0 Å, the sampling interval must be less than 1.6. In the case of a one-dimensional Heisenberg model with 10 sites, the sampling interval should not exceed 0.18 when $J = h = 1$. For $J = 1$ $h = 2$, even a sampling interval of 0.14 would cause deviations in the peak positions of the spectrum. For a 4×3 two-dimensional Heisenberg model, the threshold is even smaller. A sampling interval of 0.1 is suitable for $J = h = 1$, but not for the $J = 1$ $h = 2$ model. Additionally, Fig. 3 demonstrates that even if the sampling interval exceeds this threshold, not all spectral peaks exhibit deviations. Typically, peaks corresponding to eigen-energies with larger absolute values shift, while those corresponding to eigen-energies with smaller absolute values remain accurate.

The primary reason for this phenomenon lies in the excessively long sampling intervals. According to the Nyquist-Shannon sampling theorem, if a signal's maximum frequency is $f_{max}$, in order to reconstruct the original signal without distortion, the sampling frequency must be at least $2f_{max}$. For the $Q(t)$ signal, based on Eq. 5, its frequencies are determined by various of eigen-energies $E_i$: $f_i = \frac{|E_i|}{2\pi}$. Therefore, the maximum frequency of $Q(t)$ is determined by the eigen-energy with the largest absolute value. Generally, for any quantum system, the ground state energy often has the largest absolute value: $f_{max} = \frac{|E_0|}{2\pi}$. Given that the ground state energy of the H4 system with a bond length of 1.0 Å is -2.1664 Ha, the maximum sampling interval can be determined as:

$$\Delta_{max} = \frac{1}{2f_{max}} = \frac{\pi}{|E_0|} = 1.45 \tag{18}$$

Thus, for spectra with $\Delta = 1.5$ or 1.8, as shown in Fig. 3(a), the sampled $q(n)$ signal experiences spectral aliasing, leading to an inability to accurately reconstruct the original spectral graph. For the systems shown in Fig. 3(b)–(f), the maximum sampling intervals are 1.66, 0.184, 0.135, 0.117, and 0.081, which align with the observations in the figure. To reduce sampling cost, the algorithm can utilize the largest possible sampling interval during execution. Although the exact eigenenergy with the largest absolute value (often can be seen as the ground state energy) is unknown, classical approximation algorithms can be employed to estimate an upper bound of the ground state energy $E_0$. By substituting this upper bound into Eq. (18), a relatively optimal value for $\Delta$ can be determined. Given the approximate proportionality between the magnitude of the ground state energy and the system size $L$, such as in the case of an infinite one-dimensional Heisenberg model where the eigen-energies asymptotically scale linearly with the number of lattice sites, we find that: $\Delta \propto 1/|E_0| \propto 1/L$.

It is important to note that, in theory, based on Eq. (13), the height of a spectral peak directly reflects the overlap between the reference state and the corresponding eigenstate, i.e., $|c_i|^2 = a_i$. However, even when the sampling interval satisfies the Nyquist sampling theorem, spectral leakage can still occur due to non-periodic sampling, resulting in $|c_i|^2 \neq a_i$. While this causes quantitative inaccuracies in calculating the wavefunction overlap, the qualitative analysis remains reliable. In other words, the height of each spectral peak can still be reliably used to identify which eigenstates dominate in the reference state.

### 3.3 Determining the Sign of Eigenvalues

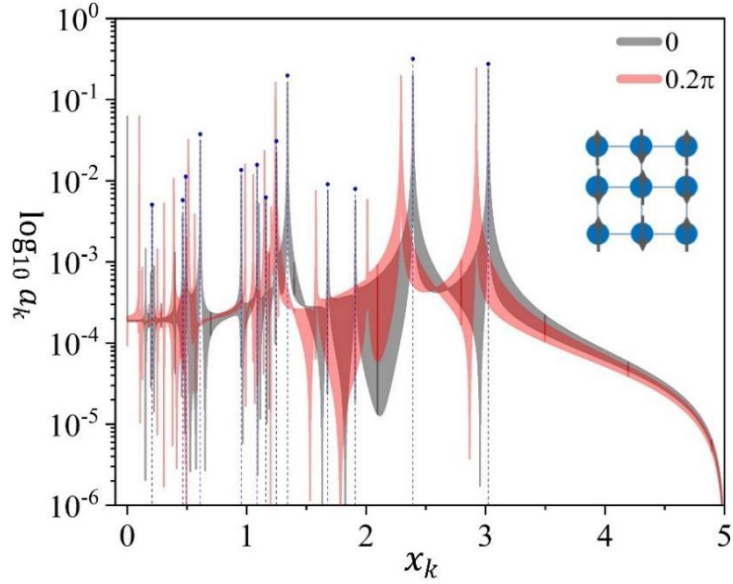

Figure 4: DFT spectra of the $q(n)$ signals for a 3×3 two-dimensional Heisenberg model with $J = h = 1$, where $T_{\max} = 1000$ and $\Delta = 0.1$ for all sampling sequences. The x-axis, y-axis, and blue dotted vertical lines have the same meaning as those in Figure 3. The only distinction between the black and red lines is that the red line has a positive constant offset of $0.2\pi$ applied to the Hamiltonian, while the black line has no offset.

It is worth noting an issue of the proposed algorithm: WQTE can only determine the absolute value of the eigen-energy, but not its sign. However, this is not a significant problem. Firstly, for the majority of quantum systems, eigen-energies are typically negative, so when calculating some low-energy eigenvalues, their absolute values can be directly taken as negative. Secondly, even if the sign cannot be determined, it can be inferred by applying a small positive constant offset $s_0$ to the Hamiltonian: the eigenvalues of the original Hamiltonian $\hat{H}$ are $E_i$, and those after the offset are $E_i + s_0$. When $E_i$ is positive, $|E_i + s_0|$ will be greater than $|E_i|$, resulting in a right shift of the spectral peak corresponding to $E_i$; when $E_i$ is negative, $|E_i + s_0|$ will be smaller than $|E_i|$, resulting in a left shift of the peak in the corresponding spectrum. Therefore, by performing the WQTE algorithm separately on the Hamiltonian $\hat{H}$ and $\hat{H} + s_0$, and comparing their spectra, the direction of the peak shift can determine the sign of the eigenvalue. Notably, the selection of $s_0$ requires careful consideration to avoid two potential issues: if $s_0$ is smaller than the spectral resolution $\delta$, the direction of peak shift becomes difficult to identify, thereby affecting the determination of eigenvalue signs; if $s_0$ is excessively large (exceeding $\pi/\Delta - |E_{max}|$), it may lead to spectral aliasing. To balance these trade-offs, $s_0$ should not be overly large—principally, $s_0$ only needs to be slightly larger than $\delta$ to ensure effective sign discrimination. For more robust identification in practical applications, we recommend an empirical range of $s_0 = 5\delta \sim 10\delta$. As shown in Fig. 4, for a 3×3 two-dimensional Heisenberg model with $J = h = 1$, the leftward shift of the peaks near 1.34, 2.4, and 3.0 along the x-axis indicates negative eigen-energies, whereas the rightward shift of the peaks near 0.95 and 1.9 suggest positive eigen-energies, both consistent with the ED results.

## 3.4 Analysis of Computational Accuracy

**Absolute Error Analysis**

The theoretical analysis indicates that the computational error is solely determined by the maximum evolution time. This is demonstrated numerically using an $H_4$ molecular chain with a bond length of 1.0 Å, as shown in Fig. 5. In most cases, the eigen-energies $E_i$ cannot be expressed as integer multiples of $\delta$, leading to quantization errors. To minimize these errors, the quantization unit $\delta$ should be as small as possible. From the analysis, $\delta = 2\pi/(N \cdot \Delta) = 2\pi/T_{max}$, implying that the longer the maximum evolution time, the higher the spectral resolution. Fig. 5(a) presents spectra of various $q(n)$ sequences obtained with $\Delta = 1.0$ under different $T_{max}$, with ED values serving as the benchmark. As shown, none of the selected $T_{max}$ values allow the ground state energy to be an exact integer multiple of $\delta$. Consequently, all spectral peaks around $x_k = 0.35$ deviate from the ground state energy. However, as $T_{max}$ increases, the deviation significantly decreases due to improved spectral resolution. Fig. 5(b) quantitatively describes the relationship between this deviation and $T_{max}$ by showing the absolute deviation between the eigenvalues from the spectral peak positions and those obtained via ED. Since the WQTE method can only resolve eigenstates with non-zero overlap with the reference state, only the errors for computable eigenvalues are shown. For some eigen-energies (e.g., $E_{14}$), the deviation remains small even when $T_{\max}$ is short (i.e., $\delta$ is large), primarily due to these eigen-energies being close to integer multiples of $\delta$, a result of fortuitous coincidence. Nevertheless, even when eigen-energies deviate from integer multiples of $\delta$, the deviation will not exceed $\delta$, making $\delta$ the upper bound for the error. From Fig. 5(b), it is evident that at a constant sampling frequency, the upper limit of computational error shows an approximately inverse relationship with $T_{\max}$. To elucidate this, we fit the maximum error with the function $y = A/x$, yielding $E_{\mathrm{err}} = 5.964/T_{\max}$. The fit exhibits a high degree of accuracy with $R^2 = 0.98695$, and the fitted coefficient of 5.964 is very close to $2\pi$, further validating the correctness of the theoretical analysis presented.

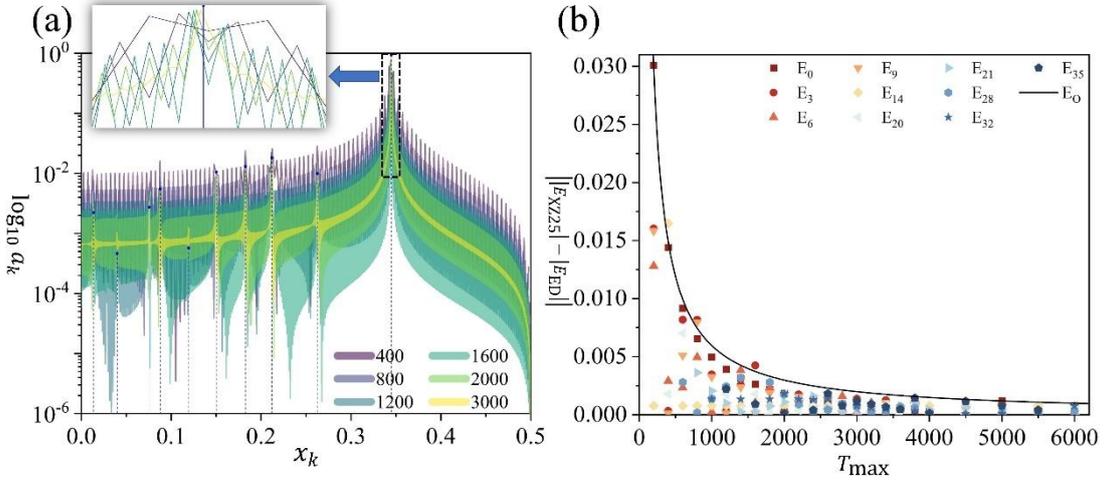

Figure 5: The WQTE calculation results for an $H_4$ molecular chain with a bond length of 1.0 Å. Panel (a) shows the spectra of $q(n)$ signals with $\Delta = 1.0$, but for different maximum evolution times ($T_{\max}$), as indicated in the legend. The x-axis, y-axis, and blue dotted vertical lines have same meanings with those in Figure 3. The figure also includes a magnified view of the spectral plot, focusing on the x-axis range between 0.34 and 0.35. Panel (b) depicts the computational errors for the $q(n)$ signals, again with $\Delta = 1.0$ but with varying $T_{\max}$. The y-axis represents the absolute difference between the eigen-energies computed by the WQTE method and those obtained via ED for each eigenstate. The x-axis shows different values of $T_{\max}$, with points corresponding to various eigenstates. The black solid line represents the fitted maximum error as a function of $T_{\max}$.

**Relative Error Analysis**

Unlike the requirement for chemical accuracy in quantum chemistry calculations, many quantum models prioritize relative accuracy, expressed as $|E_{WQTE} - E_{ED}|/|E_{ED}|$. Based on the analysis above, the absolute error $|E_{WQTE} - E_{ED}|$ is determined by $\delta = \frac{2\pi}{T_{max}} \leq |E_{WQTE} - E_{ED}|$, while the maximum sampling interval is determined by the ground state energy: $\Delta_{max} = \pi/|E_0|$. Therefore, according to Eq. (14), when $\Delta$ is set to its maximum value, the total number of sampling steps for the evolution time is:

$$N = \frac{2\pi}{\Delta_{max}\delta} \geq \frac{2|E_0|}{|E_{WQTE} - E_{ED}|} \tag{19}$$

This implies that once the maximum relative error is determined, $N$ is also fixed. According to Eq. (19), when the relative error is set to 0.1%, sampling only 2000 evolution time points, with the sampling interval at its maximum value, ensures that the calculated ground state energy meets the relative error constraint. However, since this method is also used to compute higher-energy eigenstates, and given that the absolute error remains constant while the eigen-energies $|E_i|$ for higher states are smaller, the resulting relative error will be larger for those states. Additionally, because the exact value of the ground state energy is unknown in advance, the sampling interval $\Delta$ must be set slightly smaller than $\Delta_{max}$ during the computation process. These factors all affect the final relative error. Therefore, when executing the algorithm, some redundancy in the sampling number $N$ is necessary; for instance, setting $N = 4000$ ensures that the relative error remains below 0.1%. (Of course, considering the symmetry of the $q(n)$ sequence, where $q(n) = q(N - n)$, only the sequence from $n = 0$ to $n = 2000$ needs to be sampled on the quantum processor.) To numerically demonstrate this, Fig. 6 presents the relative error obtained using the WQTE method for various Heisenberg models. In all cases, the sampling number $N$ is set to 4000, and the sampling interval $\Delta$ is slightly smaller than their respective $\Delta_{max}$.

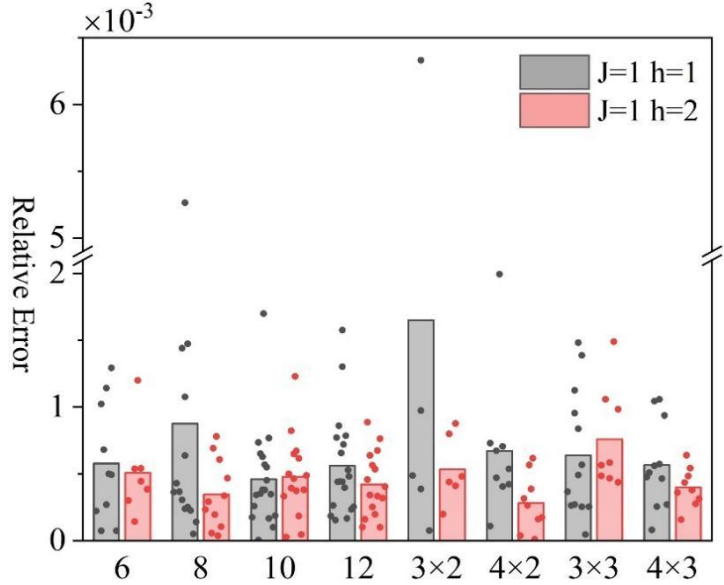

Figure 6: Relative errors in the eigen-energies of various Heisenberg models computed using the WQTE method. The X-axis represents different models, ranging from left to right: one-dimensional Heisenberg models with 6, 8, 10, and 12 lattice sites, followed by two-dimensional Heisenberg models with lattice configurations of 3×2, 4×2, 3×3, and 4×3. The Y-axis represents the relative error, $|E_{XZ25} - E_{ED}|/|E_{ED}|$, where $E_{ED}$ denotes the eigen-energies calculated using the ED method. Red and black colors represent different model parameters, as specified in the legend. For all calculations, the sampling number $N$ is set to 4000, and the sampling interval $\Delta$ are set as follows, from left to right: 0.3, 0.2, 0.2, 0.15, 0.18, 0.12, 0.15, 0.1, 0.25, 0.18, 0.16, 0.11, 0.18, 0.13, 0.1, and 0.08. Each point in the figure represents the relative error for one eigenvalue of the respective model, and the bars represent the average relative error across all eigenvalues for that model.

In Fig. 6, the relative error for most of the calculated results remains within 0.1%, and the average relative error across all models also stays below 0.1%, with the exception of the 3×2 two-dimensional Heisenberg model with parameters $J = h = 1$. The larger average relative error in this model is due to one eigenstate with an energy of 0.05059, while the WQTE method computes it as 0.05027. Although the absolute error between these values is small, the eigenvalue itself is close to zero, resulting in an inflated relative error. Similar reasons explain the relatively larger errors in other eigenvalues shown in Fig. 6. In general, we are primarily concerned with low-energy eigenstates, whose eigen-energies are not far from the ground state and tend to have larger absolute values. Therefore, the likelihood of excessively large relative errors occurring for these eigenstates is low.

## 3.5 Analysis of the Sampling Error

**Simulation Results**

All the preceding analysis and discussion are predicated on the assumption that the $q(n)$ sequence can be sampled with high accuracy. In the implementation of the WQTE algorithm, obtaining $q(n)$ requires measuring the probability of the ancillary qubit collapsing into the $|1\rangle$ and $|0\rangle$ states, with the difference between these probabilities providing the value of $q(n)$. However, due to the finite number of quantum state measurements, accurately determining the collapse probabilities is challenging, and the resulting $q(n)$ values are inherently subject to sampling errors. To evaluate the impact of these errors on the algorithm's performance, Fig. 7 presents the effect of different quantum state measurement counts $M$ on the $q(n)$ spectrum for various systems. Comparing the spectra for various measurement counts reveals that sampling error appears as white noise superimposed on the error-free $q(n)$ spectrum. Unlike the error-free $q(n)$ signal, this white noise is uniformly distributed across all frequency components, with random amplitudes and a consistent power density across the spectrum, ensuring that no single frequency is dominant. Consequently, when the number of quantum state measurements is low, eigenstates with small $a_k$ values in the $q(n)$ spectrum may be obscured by the white noise, rendering them difficult to discern. Nevertheless, the accuracy of eigenvalue estimates for more prominent peaks (those with larger $a_k$ values) remains unaffected. As the number of measurements $M$ increases, the overall amplitude of the white noise decreases, allowing lower peaks to become distinguishable, and the associated eigenvalues can be obtained via Eq. (14).

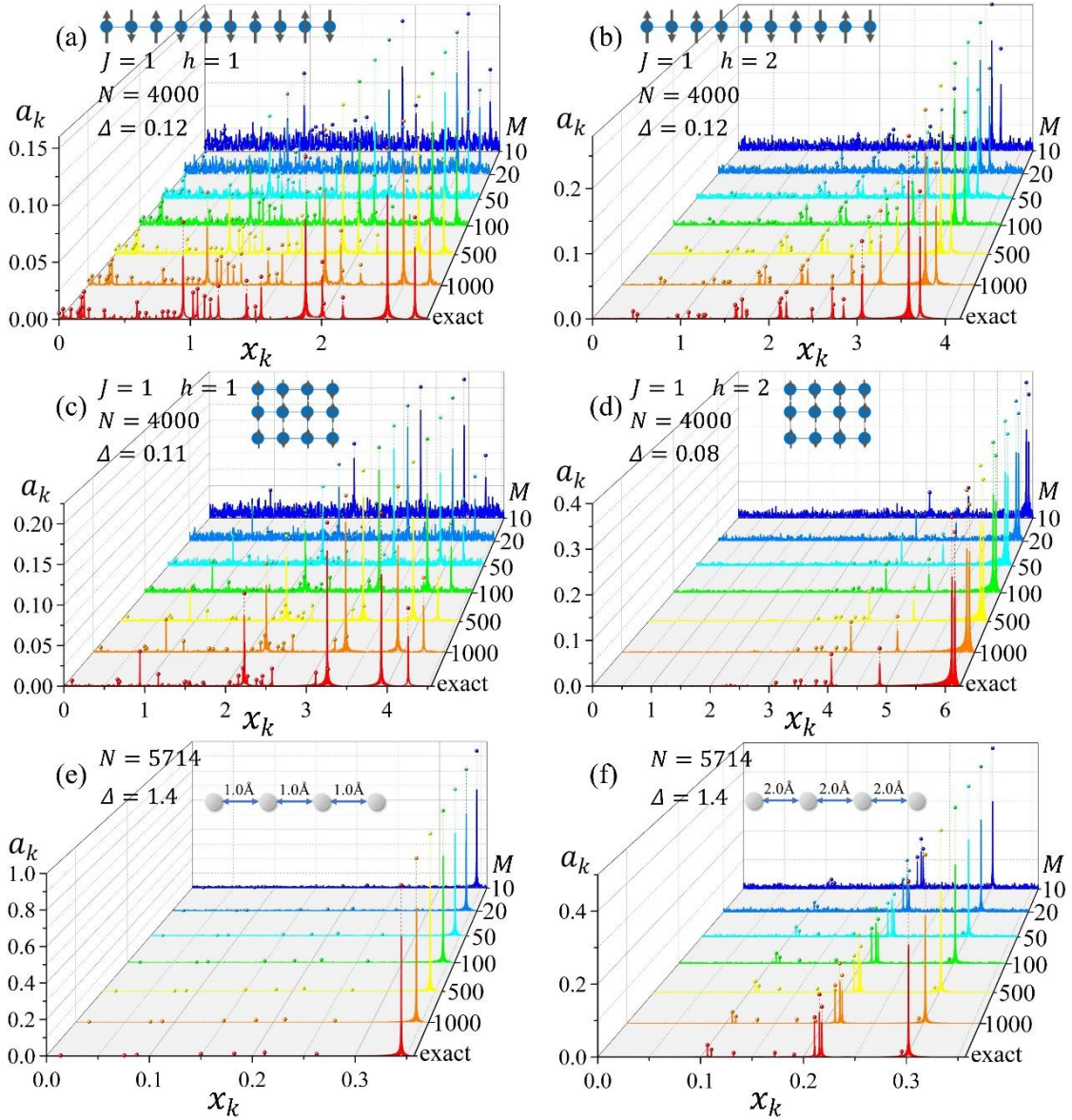

Figure 7: The spectral plots of $q(n)$ sequences estimated under different quantum circuit sampling numbers (M=10,20,50,100,200,500,1000, and 'exact') for various Heisenberg models and the H$_4$ molecular chain. The configurations of the different models, the reference states selected for the calculations, and other relevant computational parameters are annotated in the figure. The $x_k$, $a_k$ and the dotted vertical lines retain the same meanings as in Fig. 3. The different colored spectral lines represent different values of $M$, with 'exact' corresponding to an infinite circuit sampling number, indicating the results without sampling errors.

As shown in Fig. 7(a), when $M = 10$, the small number of samples results in significant white noise interference with high amplitude in the spectrum. However, distinct peaks at 0.944, 1.877, 2.502 and 2.71 in X-axis still stand out clearly from the white noise, allowing the corresponding eigenstates to be identified. Although there is also a peak at $x_k = 2.006$, its height is close to the white noise amplitude, making it difficult to confidently determine whether it represents a genuine eigenvalue or a random peak caused by sampling error. Other peaks with smaller $a_k$ values are completely buried in the white noise from sampling error. When $M = 50$, the peak at $x_k = 2.006$ shows a clear distinction from the white noise, enabling identification of the corresponding eigenstate. Additional peaks appear at 0.190, 1.021, 1.427, and 1.540 in $x_k$ positions, which can also be confirmed. At $M = 1000$, the white noise amplitude becomes very small, and most peaks corresponding to eigenstates become visible. Furthermore, all peak positions match exactly with the peaks in the error-free $q(n)$ spectrum. Similar phenomena are observed in calculations for other systems. In Fig. 7(b), only three eigenvalues can be identified when $M = 10$, seven eigenvalues at $M = 100$, and fourteen eigenvalues at $M = 1000$. Fig. 7(d) reveals that while $M = 10$ suffices to identify the four eigenstates with strong reference state overlap, detecting the weakly coupled eigenstates requires $M > 500$ due to their diminished spectral weights. In Fig. 6(g), since the reference state has a near-perfect overlap with the ground state but much smaller overlap with other eigenstates, the ground state is easily identified, but detecting other eigenstates requires significantly increasing the number of measurements.

In conclusion, while the number of quantum state measurements does not impact the accuracy of the computational results, sampling errors caused by a finite number of measurements introduce white noise across the entire frequency spectrum. This noise, with random amplitudes at individual frequency points but uniformly distributed across the band, can obscure eigenstates with smaller peak values, thereby reducing the algorithm's efficiency in identifying eigenstates. When the overlap between the chosen reference state and an eigenstate's wavefunction is large, fewer quantum measurements are sufficient to calculate the corresponding eigenenergy. Conversely, when the overlap is small, increasing the number of measurements becomes necessary.

**Theoretical Analysis**

The preceding discussion qualitatively outlines the influence of the quantum measurement count on the performance of the WQTE algorithm. To further this analysis, we now introduce a quantitative framework by developing a statistical model that accounts for sampling errors and explores the spectral distribution of these errors. Specifically, we aim to determine the upper bound of the peak height in the noise spectrum resulting from sampling errors when the sampling number of the evolution time is $N$, and the number of quantum measurements is $M$. If the observed peak exceeds this theoretical upper bound, it can be inferred that the peak is associated with an eigenstate, thus providing a theoretical basis for identifying spectral peaks. In the following derivation, we denote the sampling error as $s(n)$, and the estimate of $q(n)$, which incorporates the sampling error, as $\bar{q}(n)$. This relationship can be expressed as follows:

$$\bar{q}(n) = q(n) + s(n) \tag{20}$$

For a qubit, the probabilities of collapsing to the $|0\rangle$ and $|1\rangle$ states are denoted as $P_0$ and $P_1$, respectively. Through calculation, the expected value of $\bar{q}(n)$ for a single measurement is $E[\bar{q}(n)] = q(n) = P_1 - P_0$, and its variance is $D[\bar{q}(n)] = 4P_0P_1$. This indicates that the sampling variance is inherently linked to the state of the qubit. Since $0 \leq 4P_0P_1 \leq 1$, we simplify the subsequent analysis by assuming the worst-case scenario, i.e., $D[\bar{q}(n)] = 1$. According to the law of large numbers, when the quantum state is measured $M$ times, $\bar{q}(n)$ follows a normal distribution with an expectation of $q(n)$ and a variance of $1/M$. Substituting this result into Eq. (20), we find that the random signal $s(n)$ follows a normal distribution $N(0, 1/M)$. To reduce sampling costs, the symmetry $\bar{q}(n) = \bar{q}(-n)$ is utilized when sampling $\bar{q}(n)$ sequences at different $n$ values during algorithm execution. This symmetry implies that $s(n)$ also satisfies $s(n) = s(-n)$. By analogy with Eq. (12), the forward and inverse Fourier transforms of $s(n)$ are given by:

$$\begin{aligned} S(k) &= \sum_{n=0}^{N-1} s(n) e^{-i\frac{2\pi}{N}kn} = s(0) + 2\sum_{n=1}^{\frac{N-1}{2}} s(n) \cos\left(\frac{2\pi}{N}nk\right) \\ s(n) &= \frac{1}{N}\sum_{k=0}^{N-1} S(k) e^{i\frac{2\pi}{N}nk} = b_0 + \sum_{k=1}^{\frac{N-1}{2}} b_k \cos\left(\frac{2\pi}{N}nk\right) \end{aligned} \tag{21}$$

where $b_0 = S(0)/N$ and $b_k = 2S(k)/N$. It can be observed that $S(k)$ is a linear combination of the $N$-term sequence $s(n)$. Since each $s(n)$ is independent and follows the $N(0, 1/M)$ distribution, by the properties of the normal distribution, $S(k)$

also follows a normal distribution, with the expectation and variance given by:

$$E[S(k)] = 0$$
$$D[S(k)] = \frac{1}{M} + \frac{1}{M} \cdot 4 \sum_{n=1}^{\frac{N-1}{2}} \cos^2\left(\frac{2\pi}{N}nk\right) \tag{22}$$

For the term $\sum_{n=1}^{(N-1)/2} \cos^2\left(\frac{2\pi}{N}nk\right)$, when $k = 1$, we have:

$$\sum_{n=1}^{\frac{N-1}{2}} \cos^2\left(\frac{2\pi}{N}n\right) = \sum_{n=1}^{\frac{N-1}{4}} \cos^2\left(\frac{2\pi}{N}n\right) + \sum_{n=\frac{N+3}{4}}^{\frac{N-1}{2}} \cos^2\left(\frac{2\pi}{N}n\right)$$
$$= \sum_{n=1}^{\frac{N-1}{4}} \left[\cos^2\left(\frac{2\pi}{N}n\right) + \cos^2\left(\frac{2\pi}{N}\left(n + \frac{N-1}{4}\right)\right)\right] \tag{23}$$

Using $\cos\left(\theta + \frac{\pi}{2}\right) = -\sin(\theta)$, we obtain:

$$\cos\left(\frac{2\pi}{N}\left(n + \frac{N-1}{4}\right)\right) = -\sin\left(\frac{2\pi}{N}n - \frac{\pi}{2N}\right) \tag{24}$$

When $N$ is large, $\frac{\pi}{2N} \to 0$, and we can approximate the $\cos^2\left(\frac{2\pi}{N}\left(n + \frac{N-1}{4}\right)\right)$ term using the first-order Taylor expansion:

$$\cos^2\left(\frac{2\pi}{N}\left(n + \frac{N-1}{4}\right)\right) \approx \left[\sin\left(\frac{2\pi}{N}n\right) - \frac{\pi}{2N} \cdot \cos\left(\frac{2\pi}{N}n\right)\right]^2$$
$$\approx \sin^2\left(\frac{2\pi}{N}n\right) - \frac{\pi}{2N} \cdot \sin\left(\frac{4\pi}{N}n\right) \tag{25}$$

Substituting this into Eq. (23), we get:

$$\sum_{n=1}^{\frac{N-1}{2}} \cos^2\left(\frac{2\pi}{N}n\right) = \sum_{n=1}^{\frac{N-1}{4}} \left[\cos^2\left(\frac{2\pi}{N}n\right) + \sin^2\left(\frac{2\pi}{N}n\right) - \frac{\pi}{2N} \cdot \sin\left(\frac{4\pi}{N}n\right)\right]$$
$$= \frac{N-1}{4} - \frac{\pi}{2N} \cdot \sum_{n=1}^{\frac{N-1}{4}} \sin\left(\frac{4\pi}{N}n\right) = \frac{N-1}{4} \tag{26}$$

The above derivation relies on the conclusion that the $4\pi n/N$ represents angles that are uniformly distributed on the unit circle. As $N$ is large, the sine values of these angles tend to cancel each other out. Substituting this into Eq. (22), we have $D[S(1)] = N/M$. Furthermore, for any frequency $k$, we can similarly derive the variance of $S(k)$, expressed as follows:

$$D[S(k)] = \frac{N}{M} \tag{27}$$

And given that $b_k = 2S(k)/N$, we get:

$$E[b_k] = 0 \qquad D[b_k] = \frac{4}{NM} \tag{28}$$

Considering that the amplitude at each frequency in the spectrum is taken as the absolute value of the Fourier transformed result, the signal $s(n)$ will be represented as $|b_k|$ form in the frequency-amplitude spectrum. Therefore, the sampling error will be superimposed onto the spectrum of $q(n)$ in the form of a half-normal distribution: $|b_k| \sim \text{Half Normal}\left(0, \frac{4}{NM}\right)$. The scale parameter for $|b_k|$, which corresponds to the standard deviation of the normal distribution, is $\sigma = \sqrt{\frac{4}{NM}}$. Based on the properties of the half-normal distribution, we know that the probability of $|b_k|$ falling within $4\sigma$ is 0.999937. This implies that if a spectral peak with an amplitude greater than $4\sigma$ is observed, the probability that it originates from the random noise caused by sampling error is only 0.0063% (i.e., 1 in 16,000). Thus, it can be concluded with high confidence that such a peak originates from the spectrum of $q(n)$ without sampling error, i.e. from the eigenstate.

**Comparing with VQE method**

The above analysis provides a reliable criterion for peak identification: based on the number of evolution time samples, $N$, and the number of quantum state measurements, $M$, a threshold parameter $4\sigma$ can be established to filter the spectral peaks. Only peaks with heights exceeding this threshold are accepted, and the corresponding eigenstates can be computed using Eq. (14). As demonstrated, when $N$ is sufficiently large, if there is significant overlap between the eigenstate to be determined and the reference state, sampling noise will barely obscure these eigenstates, even when the quantum state measurement count $M$ is small. For instance, in the calculations shown in Fig. 7(a)~(d), where the evolution time sample number is $N = 4000$, if we are only interested in eigenstates with overlap $|c_i|^2 > 0.1$, measuring the quantum state just twice (i.e., $M = 2$, where the possible values of $\bar{q}(n)$ are limited to -1, 0, and 1) is sufficient, as the threshold $4\sigma \approx 0.09$ under this condition. To verify our analysis, Fig. 8(a)~(d) replicate the calculations from Fig. 7(a)~(d), except that the number of quantum state measurements is set to 2 in all cases in Fig. 7. From the figures, it is clear that all peaks exceeding the $4\sigma$ threshold correspond to eigenstates. It is worth noting that some eigenstates inherently satisfy $|c_i|^2 > 4\sigma$, but due to spectral leakage, their amplitudes

$a_i < 4\sigma < |c_i|^2$, leading to the failure to identify these eigenstates.

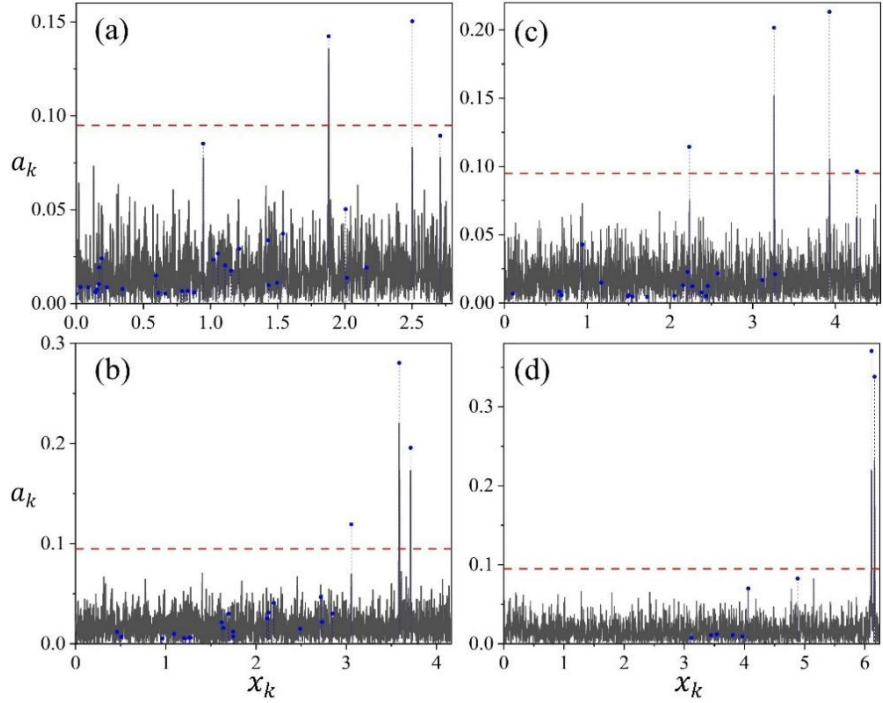

Figures 8(a)~(d) correspond to the spectra of $q(n)$ shown in Figures 7(a)~(d), where the number of quantum measurements is set to $M = 2$. The blue dotted vertical lines have the same meaning as those in Figure 3. The horizontal red dashed line represents the threshold height of $4\sigma \approx 0.09$.

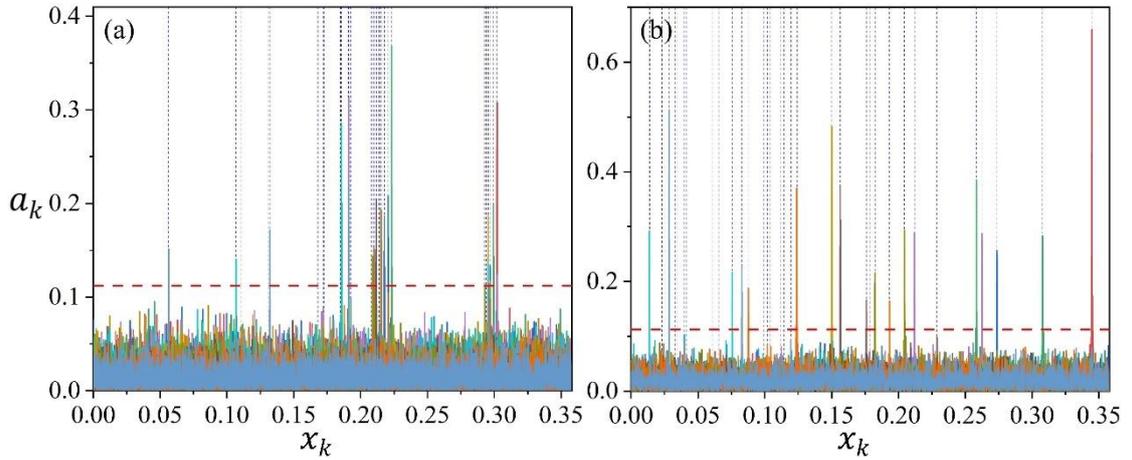

Figure 9(a)(b) corresponds to the $q(n)$ spectrum in Fig. 7(e)(f), where the number of quantum measurements is set to 1. The curves in different colors represent the $q(n)$ spectrum calculated under different reference states. The blue vertical dashed lines indicate the positions of all eigenstates, while the red horizontal dashed line represents the threshold height of $4\sigma \approx 0.11$.

To identify eigenstates with smaller $|c_i|^2$, a more efficient approach than increasing the number of quantum measurements is to select a different reference state. For example, in the case of the H$_4$ molecular chain with the equilibrium structure shown in

Fig. 7(e), the reference state is the HF state, with $N = 5714$. Even with only one quantum measurement ($M = 1$, where the possible values of $\bar{q}(n)$ are limited to -1 and 1), the threshold is $4\sigma \approx 0.11$. Since the ground state component overwhelmingly dominates in the HF state, the overlap between the two wavefunctions is nearly 1, making the ground state easily detectable—though only the ground state is obtained. To compute some of the lower-lying excited states, one can switch the reference state to a single or double excitation configuration, as these states typically have substantial overlap with low-energy excited states without significantly increasing the quantum circuit cost. The results are shown in Fig. 9(a), where it is evident that, even with just one quantum measurement, most of the low-energy states can be identified as long as an appropriate reference state is selected. For the dissociated $H_4$ configurations, as shown in Fig. 9(b), our results demonstrate that even for strongly correlated systems, the proposed method remains effective in computing various eigen-energies through testing different reference states, without significant computational cost escalation. In contrast, conventional quantum chemical methods struggle to accurately determine these eigen-energies. Even the widely studied VQE approach faces considerable challenges in constructing appropriate ansatz for strongly correlated systems - let alone designing ansatz circuits capable of preparing excited states.

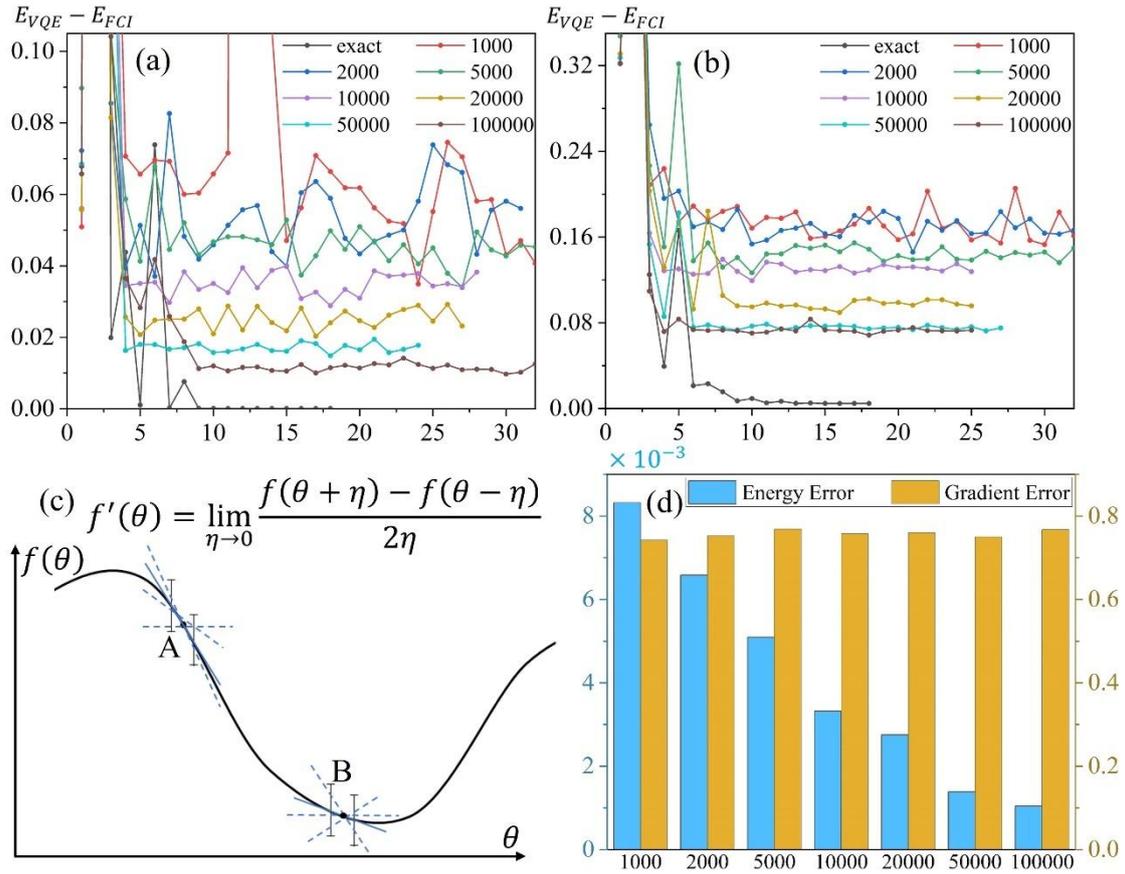

Figure 10. Convergence behavior of the UCCSD-VQE method for H$_4$ molecular chains with bond lengths of 1.0 Å (a) and 2.0 Å (b) under different sampling numbers. The horizontal axis represents iteration steps, while the vertical axis shows the energy difference between the obtained energy at each step and the FCI ground-state energy. Curves with different colors correspond to varying sample sizes used for measuring expectation values of each Pauli string in the Hamiltonian. (c) illustrates the impact of sampling errors on gradient calculations. (d) displays both energy deviations and gradient deviations during VQE iterations with different sampling numbers (obtained by averaging the differences between sampled results and their exact values).

To compare with the VQE method in terms of sampling cost, we implemented VQE with varying sample sizes for the same H$_4$ system using a UCCSD ansatz. As shown in Figs. 10(a) and 10(b), under ideal conditions with zero sampling error, the VQE method converges accurately to the ground state for both equilibrium and dissociated configurations. However, in practical VQE implementations, estimating the Hamiltonian expectation value of the output state requires multiple circuit samplings, where finite sampling inevitably introduces measurement errors. Consequently, even when the optimized circuit produces the exact ground state, the measured energy still deviates from the true ground state energy. More critically, when computing parameter gradients via the finite-difference method (as demonstrated for point A in Fig. 10(c)), sampling errors perturb the values of $f(\theta \pm \eta)$, leading to inaccurate gradient estimations. This effect becomes particularly pronounced near extremal points where gradients approach zero - the difference between $f(\theta + \eta)$ and $f(\theta - \eta)$ becomes dominated by random fluctuations from sampling noise rather than true energy variations. This explains the optimization behavior observed in Figs. 10(a) and 10(b): while VQE rapidly converges near the ground state within 10 iterations, subsequent refinement becomes increasingly difficult, manifesting as persistent energy oscillations. At this stage, gradient-based optimization effectively stalls due to unreliable derivative information. Although increasing measurement counts improves results, non-negligible deviations from the exact ground state energy persist. Fig. 10(d) quantifies this by plotting the average energy and gradient deviations (differences between sampled and exact values) during VQE iterations across different sample sizes. While energy deviations decrease with more samples, gradient deviations remain substantial and are far less sensitive to sample size compared with energy deviations. This fundamentally limits optimization precision and explains the observed oscillation plateau in energy convergence curves.

It should be noted that the sampling numbers mentioned in Fig. 10 do not represent the total sampling counts, but rather the number of measurements allocated for estimating the expectation value of each Pauli string in the Hamiltonian. Considering that the $H_4$ Hamiltonian contains 184 Pauli strings, the total measurements required for one Hamiltonian expectation evaluation should be multiplied by 184. Furthermore, since the ansatz circuit involves 14 parameters, calculating their gradients necessitates additional energy evaluations at slightly shifted parameter values ($\theta \pm \eta$), requiring 29 energy measurements per iteration step. The total sampling count for the VQE method is then obtained by multiplying this per-step measurement count by the number of iterations. As shown in Table 1, which compares the sampling costs and result accuracy between our method and VQE for the $H_4$ system, the WQTE method demonstrates a 4-6 orders of magnitude reduction in sampling requirements while achieving 1-3 orders of magnitude higher accuracy than VQE. These results clearly indicate the remarkable robustness of our new method against sampling errors and its absolute advantage over

VQE in terms of sampling efficiency.

Table 1. Computational accuracy comparison of the WQTE method versus VQE at varying sampling costs ($N_{steps}$: number of optimization iterations).

| Method | Total Sampling Costs | Calculation Error | |
|---|---|---|---|
| | | $H_4$(1.0 Å) | $H_4$(2.0 Å) |
| WQTE | 5714 | 6.3×10⁻⁴ | 6.2×10⁻⁴ |
| VQE | $N_{steps}$×5.3×10⁶ | 0.035 | 0.153 |
| | $N_{steps}$×1.06×10⁷ | 0.040 | 0.146 |
| | $N_{steps}$×2.65×10⁷ | 0.034 | 0.127 |
| | $N_{steps}$×5.3×10⁷ | 0.029 | 0.119 |
| | $N_{steps}$×1.06×10⁸ | 0.020 | 0.089 |
| | $N_{steps}$×2.65×10⁸ | 0.015 | 0.072 |
| | $N_{steps}$×5.3×10⁸ | 0.009 | 0.068 |

**3.6 Analysis of the compilation errors**

**Simulation of Compilation Errors**

As shown in Fig. 1, the circuit cost of the proposed algorithm primarily stems from the $e^{-i\hat{H}t}$ operator. Using second-order Trotter decomposition[40], the original operator can be expressed as:

$$e^{-i\hat{H}t} = \left(e^{-\frac{i\hat{h}_1\tau}{2}}e^{-\frac{i\hat{h}_2\tau}{2}}\cdots e^{-i\hat{h}_Z\tau}\cdots e^{-\frac{i\hat{h}_2\tau}{2}}e^{-\frac{i\hat{h}_1\tau}{2}}\right)^{\frac{t}{\tau}} + O(t\cdot\tau^2) \quad (29)$$

where $O(t\cdot\tau^2)$ represents the compilation error. To ensure compilation accuracy, we divide the total evolution time $t$ into smaller segments of duration $\tau$ (referred to as the "trotter step size" in this work), with each $e^{-i\hat{H}\tau}$ operator undergoing separate Trotter decomposition. This requires performing $t/\tau$ second-order Trotter expansions in total. A smaller $\tau$ leads to reduced compilation errors but increased circuit complexity, necessitating careful balancing between these factors. Therefore, it is crucial to investigate how compilation errors at different $\tau$ values affect algorithm performance. Fig. 11 presents simulation results for the Heisenberg model at various $\tau$ values (0.01, 0.04, 0.08, 0.12, 0.15, 0.18, 0.2), using identical computational parameters: $J = 1$  $h = 2$  $N = 4000$  $\Delta = 0.1$  $t = 0, \Delta, 2\Delta, \cdots, N\Delta$.

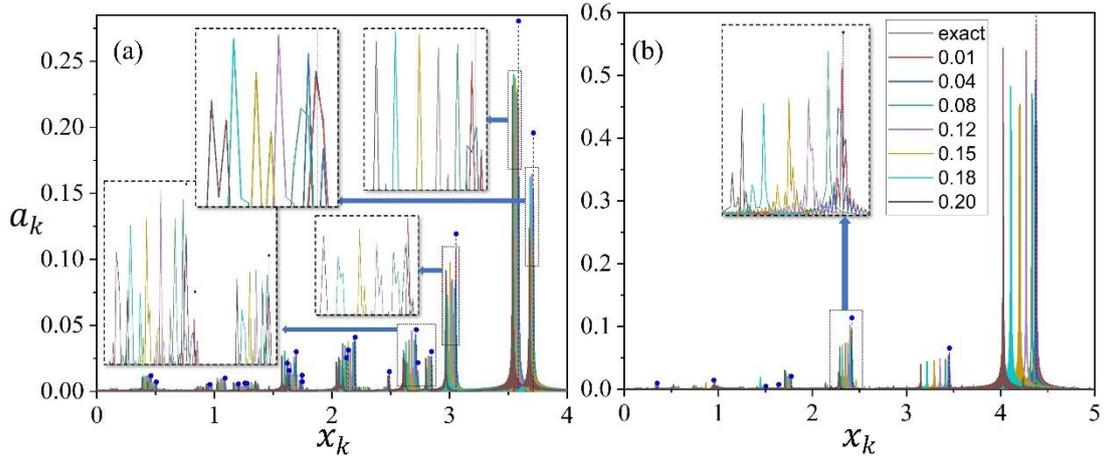

Figure 11. $q(n)$ spectra for (a) 1D 10-site and (b) 2D 3×3 Heisenberg models at different $\tau$ values. The x-axis, y-axis, and blue dotted vertical lines maintain the same meaning as in Fig. 3. Colored curves represent different $\tau$ values (see legend in panel b), with "exact" denoting spectra from precisely compiled time-evolution operators.

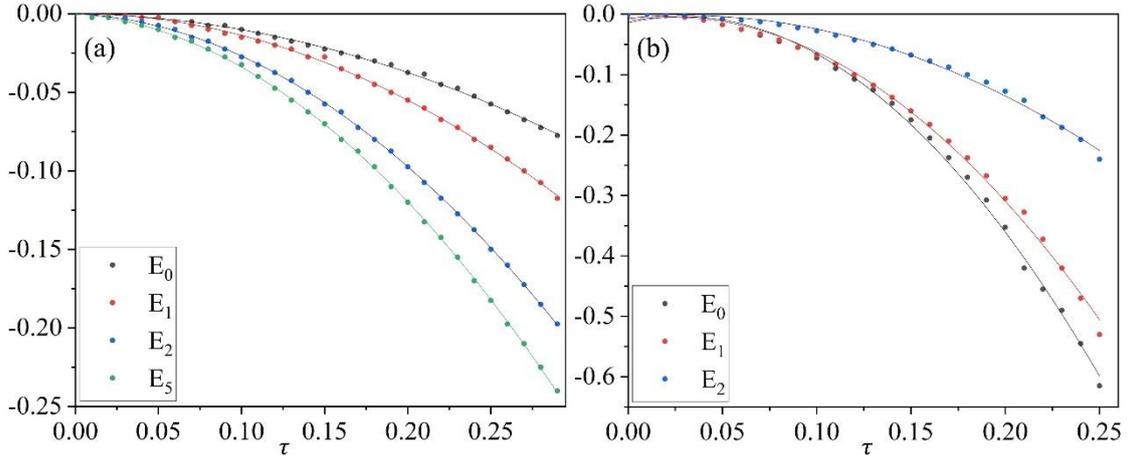

Figure 12. (a)(b) Peak shift magnitudes versus $\tau$ for different eigenstates in Figs. 11(a)(b), respectively, with curves showing quadratic fits to the data points.

The spectra maintain consistent overall features across different $\tau$ values, with all spectral peaks remaining identifiable but exhibiting position shifts that grow with increasing $\tau$. To quantify this effect, Fig. 12 plots the $\tau$-dependent peak shifts relative to the exact solution. Notably, even for fixed $\tau$, different eigenstates exhibit varying shift magnitudes, though all follow similar quadratic trends ($y = ax^2$). Quadratic fitting (solid curves in Fig. 12) achieves excellent agreement ($R^2 > 0.995$), confirming the $O(t \cdot \tau^2)$ relationship between compilation error and $\tau$-consistent with theoretical expectations from Suzuki-Trotter decomposition.

**Circuit Complexity Analysis**

Building on these results, we analyze how the circuit cost scales with target accuracy $\varepsilon$ and system size $L$. Let $N_P$ denote the total number of Pauli strings. Implementing a controlled $e^{-i\hat{h}_i\tau}$ operator requires $O(L)$ circuit complexity, making the cost for each $e^{-i\hat{H}\tau}$ operator $O(N_P \cdot L)$. Following Section 3.4's theory, when requiring energy precision $\varepsilon$, the energy quantization unit $\delta$ must satisfy $\delta \leq \varepsilon$. From Eq. 16, this dictates a minimum evolution time $T_{\max} \geq 2\pi/\varepsilon$. To ensure compilation errors $O(T_{\max} \cdot \tau^2) \leq \varepsilon$, the Trotter step size must satisfy $\tau \leq O(\sqrt{\varepsilon/T_{\max}})$, yielding the number of Trotter slices $T_{\max}/\tau \geq O(\sqrt{T_{\max}^3/\varepsilon}) = O\left((2\pi)^{\frac{3}{2}} \cdot \varepsilon^{-2}\right)$. Consequently, the quantum circuit depth scales as:

$$N_T \cdot \frac{T_{\max}}{\tau} \cdot O(L) = O\left(\sqrt{8\pi^3} \cdot N_P \cdot L \cdot \varepsilon^{-2}\right) \tag{30}$$

This analysis establishes the fundamental scaling relationship between circuit resources, system size, and target precision in our approach.

## 3.7 Analysis of the Quantum Circuit Noise

**Numerical Simulations**

In real quantum devices, due to the immaturity of current technologies, quantum systems are subject to various noise sources arising from quantum operations, environmental influences, and other factors[41,42]. At the current stage, the impact of noise on quantum devices cannot be neglected, making noise robustness one of the key metrics for evaluating the practicality of quantum algorithms. To this end, we simulated the performance of our newly proposed algorithm under quantum circuit noise. Given that the best achievable gate error rate for mainstream quantum computers with moderate qubit counts is currently around 0.5%, even a circuit with only 1,000 quantum gates would yield an output state fidelity of merely $0.995^{1000} \approx 0.67\%$. Therefore, under current gate error rates, the number of quantum logic gates should be kept to a minimum. Accordingly, we conducted a test using a random quantum circuit, as shown in Figure 13(a), which can be represented as the real-time evolution operator $e^{-i\hat{H}t}$ for some unknown Hamiltonian. For the WQTE method, we set $N = 10000$ and $\Delta = 0.4$ and only one X gate applied on Q0 qubit to prepare the reference state $|000001\rangle$. Figure 13(b) displays the noise-free DFT spectrum of the corresponding $q(n)$ signal.

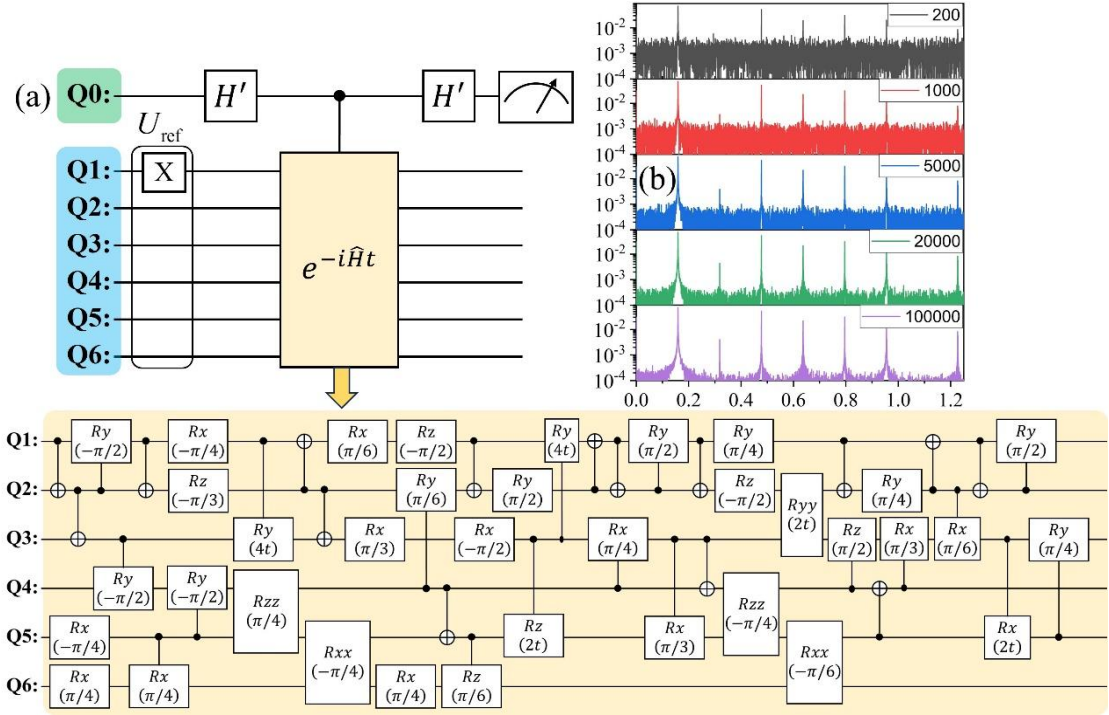

Figure 13. (a) Quantum circuits for testing WQTE method under circuit noise. (b) The corresponding DFT spectrum of q(n) under noise-free conditions for different numbers of quantum measurements.

We employed the Pauli channel model to simulate circuit noise, characterized by the probabilistic application of an additional random Pauli gate after each quantum operation[43,44]. Its mathematical representation is:

$$\varepsilon(\rho) = (1 - \gamma)\rho + \frac{\gamma}{3}(X\rho X + Y\rho Y + Z\rho Z) \tag{31}$$

Here, $\rho$ denotes the density matrix of the original quantum state, $\varepsilon(\rho)$ represents the density matrix after noise application, $\gamma$ is the noise strength parameter ranging from 0 to 1. A larger $\gamma$ indicates stronger noise; $\gamma = 0$ corresponds to the noise-free case $\varepsilon(\rho) = \rho$, while $\gamma = 1$ represents the maximum noise scenario where every gate is followed by a random Pauli operation (X, Y, or Z with equal probability).

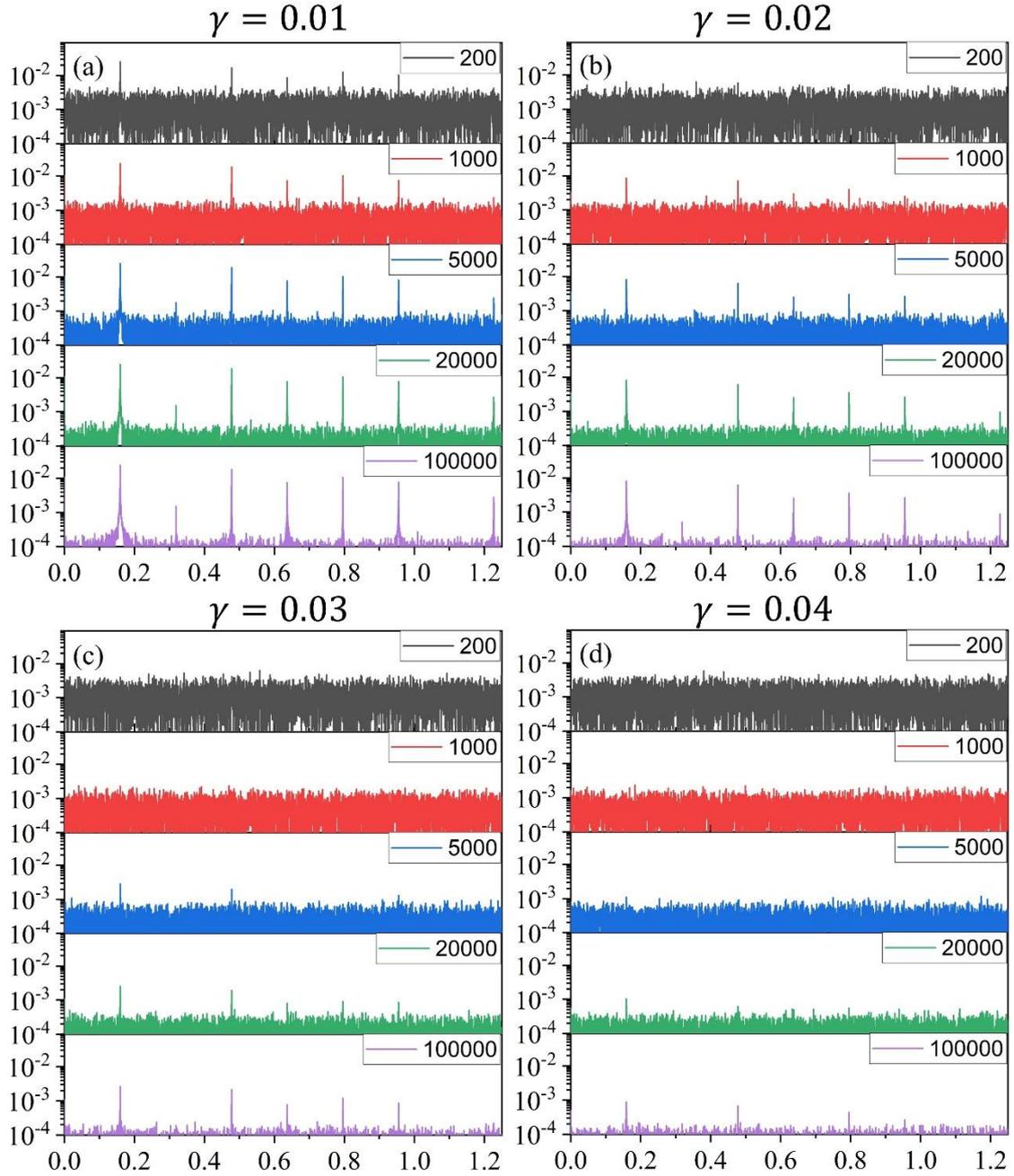

Figure 14 shows the DFT spectra of $q(n)$ obtained by applying the WQTE algorithm to the quantum circuit depicted in Figures 13(a), under different noise strengths ($\gamma = 0.01, 0.02, 0.03, 0.04$) and varying numbers of measurements per time step ($M = 200, 1000, 5000, 20000, 100000$)..

Figure 14 presents the $q(n)$ spectra obtained by executing the WQTE algorithm on the circuit in Figure 13(a) under different noise strengths ($\gamma = 0.01, 0.02, 0.03, 0.04$) and varying numbers of measurements per time step. Compared to Figure 13(b), circuit noise does not shift the spectral peak positions, indicating that it does not compromise eigenvalue accuracy—similar to the effect of sampling error analyzed earlier. However, unlike sampling noise, circuit noise uniformly suppresses the amplitudes of all characteristic peaks. As the noise coefficient $\gamma$ increases, this suppression becomes more pronounced. For instance, the highest peak (at $x_k = 0.159$) has an amplitude of approximately 0.07 in the noise-free case. For $\gamma = 0.01\sim0.04$, the corresponding peak amplitudes reduce to about 0.025, 0.008, 0.0027, and 0.001, respectively. Thus, while circuit noise affects the quantitative estimation of wavefunction overlaps between eigenstates and the reference state, Figure 14 shows that the relative heights of different peaks remain largely consistent across noise levels. Therefore, peak amplitudes retain qualitative significance for identifying dominant eigenstates.

Furthermore, comparing spectra with the same $M$ but different $\gamma$ reveals that noise strength does not significantly alter the white noise background intensity; the overall noise power depends primarily on the measurement count $M$. Circuit noise predominantly attenuates the signal strength, which may obscure weaker spectral peaks within the noise floor, but does not alter their positions or relative amplitudes. By increasing the measurement count to suppress the background noise, even attenuated peaks can be recovered.

We conclude that the WQTE algorithm exhibits inherent resilience to circuit noise: while noise attenuates spectral peak amplitudes, making them harder to detect, it does not introduce systematic errors in eigenvalue determination. Sufficiently increasing the measurement count can compensate for noise-induced attenuation, enabling reliable peak identification even in noisy environments.

**Theoretical Analysis**

We now provide a theoretical analysis of how circuit noise influences the $q(n)$ spectrum, offering qualitative insights that support the above numerical observations. Given the probabilistic nature of noise, which randomly inserts additional operations after quantum gates with probability $\gamma$, let $f(\gamma)$ denote the probability that no extra random operation occurs throughout the entire circuit execution. In this case, measuring the ancillary qubit yields the exact $q(n)$ sequence. Clearly, $f(\gamma)$ depends on both $\gamma$ and the total number of quantum gates $N_G$, approximately as:

$$f(\gamma) = (1-\gamma)^{N_G} \tag{32}$$

For scenarios where noise occurs (with probability $1-f(\gamma)$), the measured outcome is denoted as $r(n)$. To simplify the analysis, we consider a noise model focused on the $e^{-i\hat{H}t}$ operator, which dominates the circuit cost. Under noise, the implemented evolution operator effectively corresponds to a random Hamiltonian $\hat{H}'$ evolving as $e^{-i\hat{H}'t}$. Due to the randomness of noise, the eigenvalues and eigenstates of $\hat{H}'$ are also random, varying in each circuit sampling. Consequently, the probabilities for the ancillary qubit to collapse into $|0\rangle$ and $|1\rangle$ states differ each time. However, owing to the unbiased nature of the noise, we expect the overall expectation $E[r(n)] = 0$, with variance $D[r(n)] = \sigma_r^2$ where $0 \leq \sigma_r^2 \leq 1$.

Let $\bar{\bar{q}}(n)$ represent the result after $M$ circuit samplings per evolution time point. We have:

$$\bar{\bar{q}}(n) = f(\gamma)\bar{q}(n) + [1 - f(\gamma)]\bar{r}(n) \tag{33}$$

Here, $\bar{q}(n)$ and $\bar{r}(n)$ denote the sampled results of $q(n)$ and $r(n)$, respectively. Following the analysis in Section 3.5, we have:

$$\bar{q}(n) \sim N\left(q(n), \frac{1}{M}\right) \tag{34}$$

$$E[\bar{r}(n)] = 0 \quad D[\bar{r}(n)] = \frac{\sigma_r^2}{M} \tag{35}$$

Due to the linearity of expectation and the quadratic nature of variance, the expectation and variance of $\bar{\bar{q}}(n)$ can be estimated as:

$$E[\bar{\bar{q}}(n)] = f(\gamma)q(n) \tag{36}$$

$$D[\bar{\bar{q}}(n)] = f^2(\gamma)\frac{1}{M} + [1-f(\gamma)]^2 \frac{\sigma_r^2}{M} \tag{37}$$

From these expressions, we conclude that in the asymptotic limit of infinite sampling, the $\bar{\bar{q}}(n)$ signal reduces to a scaled version of the original $q(n)$ by factor $f(\gamma)$. Thus, the spectral structure remains that of the original $q(n)$ spectrum, merely attenuated by $f(\gamma)$. Since $f(\gamma)$ decreases as $\gamma$ increases, this explains the observed reduction in spectral peak amplitudes with stronger noise in Figure 14. Under finite sampling, additional white noise is superimposed on the spectrum, with intensity inversely proportional to $M$. Given that $0 < f(\gamma) \leq 1$, the variance $D[\bar{\bar{q}}(n)]$ ranges between

$1/M$ and $\sigma_r^2/M$, implying it does not grow indefinitely with noise strength. Therefore, the relationship between variance and the noise factor $\gamma$ is relatively weak, consistent with the observation in Figure 14 that the white noise amplitude remains similar across different $\gamma$ values and depends primarily on $M$.

The theoretical results confirm that circuit noise attenuates peak intensities, making them more susceptible to being masked by sampling-induced white noise. To detect weaker spectral features, increasing the measurement count $M$ suppresses the background noise, effectively trading sampling resources for enhanced noise resilience—in full agreement with the numerical simulations. We can qualitatively assess the sampling cost imposed by circuit noise. For a spectral peak to remain distinguishable above the noise floor, we require $D[\bar{q}(n)] \leq f(\gamma)q(n)$. Substituting equations (32) and (37) yields:

$$\frac{\sigma^2}{M} \leq (1-\gamma)^{N_G} q(n) \quad \Rightarrow \quad M = O\left[\left(\frac{1}{1-\gamma}\right)^{N_G}\right] \tag{38}$$

This indicates that, although the algorithm possesses inherent noise resilience, maintaining this resilience may incur an exponential sampling cost in $N_G$. When the number of logic gates $N_G$ is large, the required sampling cost could become prohibitive. However, this exponential scaling is moderated by the noise factor $\gamma$. As quantum hardware technology matures and $\gamma$ decreases, the tolerable circuit complexity for a given sampling budget improves significantly. For example, assuming a maximum acceptable sampling cost $M = 10000$: for $\gamma = 0.01$, the maximum allowable gate count $N_G \approx 916$; for $\gamma = 0.001$, $N_G \approx 9206$; and for $\gamma = 0.0001$, $N_G \approx 92099$. While current hardware with relatively high $\gamma$ may struggle to execute WQTE circuits with high gate costs, continued technological progress in reducing $\gamma$ will progressively enable the application of WQTE to increasingly complex systems.

### 3.8 Resource Analysis for the WQTE Method

In this section, we present a comprehensive analysis of the quantum resource requirements for the WQTE method based on the preceding discussions. To compute a system of size $L$ with an absolute accuracy requirement of $\varepsilon$, we systematically evaluate the following parameters: the number of qubits, quantum gate count, total sampling number, maximal evolution time, total evolution duration, and time discretization interval, among other parameters. A detailed breakdown of these computational costs is provided in Table 3, which summarizes the scaling relationships for each resource metric in terms of $L$, $\varepsilon$, and the wavefunction overlap $p = |\langle \Psi_i | \psi_{\text{ref}} \rangle|^2$.

Table 3. Computational cost of the WQTE algorithm for a system of size $L$, with absolute error $\varepsilon$ and wavefunction overlap $p$.

| | |
|---|---|
| Number of qubits | $O(L)$ |
| Maximum evolution time | $T_{\max} = \dfrac{2\pi}{\varepsilon}$ |
| Trotter step size | $\tau = O(\sqrt{\varepsilon/T_{\max}}) = O\left(\dfrac{\varepsilon}{\sqrt{2\pi}}\right)$ |
| Number of Trotter slices | $\dfrac{T_{\max}}{\tau} = O\left((2\pi)^{\frac{3}{2}} \cdot \varepsilon^{-2}\right)$ |
| Number of Pauli strings in Hamiltonian | Lattice systems: $N_P = O(L) \sim O(L^3)$<br>Molecular systems: $N_P = O(L^4)$ |
| Quantum gate count (Trotter decomposition) | $N_P \cdot \dfrac{T_{\max}}{\tau} \cdot O(L) = O\left(\sqrt{8\pi^3} \cdot N_P \cdot L \cdot \varepsilon^{-2}\right)$ |
| Maximum sampling interval | $\Delta_{\max} = \dfrac{\pi}{|E_0|} = O(\pi \cdot L^{-1})$ |
| Number of evolution time samples | $N = \dfrac{T_{\max}}{\Delta_{\max}} = \dfrac{2|E_0|}{\varepsilon} = O(2L \cdot \varepsilon^{-1})$ |
| Minimum measurements per $q(n)$ point | $M_{\min} = \dfrac{72}{Np^2} = O(32\varepsilon \cdot L^{-1} \cdot p^{-2})$ |
| Total quantum sampling number | $N_{tq} = N \cdot M_{\min} = 64 \cdot p^{-2}$ |
| Total evolution time | $\dfrac{N^2 \Delta_{\max} M_{\min}}{2} = O(64\pi \cdot \varepsilon^{-1} \cdot p^{-2})$ |

**Qubit count**: The computation requires a set of quantum registers to store the reference state, plus an ancillary qubit. For molecular systems, the required number of qubits is determined by the molecular orbitals obtained from HF calculations, which depends on the chosen basis set but generally scales linearly with the number of atoms $L$. For lattice models such as Heisenberg or Ising models, the qubit count is proportional to the number of lattice sites. Thus, the storage complexity of the WQTE method is $O(L)$.

**Circuit cost**: As discussed in Section 3.4, the maximal evolution time $T_{\max}$ is determined by the absolute precision requirement: $T_{\max} = 2\pi/\varepsilon$. Consequently, the dominant circuit cost stems from implementing the controlled $e^{-i\hat{H}T_{\max}}$ operator. When employing Trotter decomposition to compile this operator, the total evolution time $T_{\max}$ must be appropriately segmented as $\left(e^{-i\hat{H}\tau}\right)^{T_{\max}/\tau}$ to balance compilation error against circuit complexity. As analyzed in Section 3.6, to maintain the precision within $\varepsilon$, the Trotter step size must satisfy $\tau \leq O(\sqrt{\varepsilon/T_{\max}})$, resulting in slices numbers of:

$$\frac{T_{\max}}{\tau} = O\left(\sqrt{T_{\max}^3/\varepsilon}\right) = O\left((2\pi)^{\frac{3}{2}} \cdot \varepsilon^{-2}\right) \tag{39}$$

For each slice $e^{-i\hat{H}\tau}$, a second-order Trotter expansion as in Eq. (15b) is performed. The number of expansion terms depends on the count of Pauli strings $N_P$ in $\hat{H}$, whose scaling with $L$ is system-dependent: $N_P \propto L \sim L^3$ for typical lattice models, while $N_P \propto L^4$ for molecular systems. Each $e^{-i\hat{h}_i t}$ operator requires $O(L)$ gate operations, yielding a worst-case circuit complexity of:

$$N_P \cdot \frac{T_{\max}}{\tau} \cdot O(L) = O\left(\sqrt{8\pi^3} \cdot N_P \cdot L \cdot \varepsilon^{-2}\right) \tag{40}$$

It should be noted that this represents a pessimistic estimate. Practical implementations could employ adaptive Trotter step sizes or tensor network optimizations to reduce resource requirements. Future research should explore such techniques for optimizing WQTE circuits, particularly to address challenges posed by quantum decoherence and gate noise in physical devices[45–47].

**Sampling Cost**: As established in Section 3.2, the maximum sampling interval is given by:

$$\Delta_{\max} = \frac{\pi}{|E_0|} \propto \frac{1}{L} \tag{41}$$

Given that the maximal evolution time $T_{\max} = 2\pi/\varepsilon$, the required number of evolution

time samples is:

$$N = \frac{T_{max}}{\Delta_{max}} = \frac{2|E_0|}{\varepsilon} \propto \frac{L}{\varepsilon} \tag{42}$$

As discussed in Section 3.5, the sampling error does not distort the original spectral structure of $q(n)$ but introduces additive white noise superimposed on the spectrum. The noise amplitude follows a half-normal distribution across all frequency points, with scale parameter $\sigma = \sqrt{\frac{4}{NM}}$ for $M$ quantum samples per $q(n)$ signal. From the properties of the half-normal distribution, the probability that a spectral peak with amplitude $> 4\sigma$ arises purely from noise is 0.0063%. Thus, such peaks can be reliably attributed to genuine eigenstate contributions. The amplitude of an eigenstate peak in the $q(n)$ spectrum corresponds to the squared wavefunction overlap between the target eigenstate and the reference state, $p = |\langle \Psi_i | \psi_{ref} \rangle|^2$. To ensure reliable peak detection, the minimum number of samples per $q(n)$ signal, $M_{min}$, must satisfy $4\sigma < p$, leading to:

$$M_{min} = \frac{64}{Np^2} = O(32\varepsilon \cdot L^{-1} \cdot p^{-2}) \tag{43}$$

The total quantum samples required are then:

$$N_{tq} = N \cdot M_{min} = 64 \cdot p^{-2} \tag{44}$$

Remarkably, if only relative precision $\epsilon$ is required, Eq. (19) in Section 3.4 yields $N' = 2/\epsilon$, and thus:

$$M'_{min} = 32\varepsilon \cdot p^{-2} \qquad N'_{tq} = N' \cdot M'_{min} = 64 \cdot p^{-2} \tag{45}$$

This reveals that the total sampling count is independent of system size, depending solely on the overlap $p$. As $p$ increases, $N_{tq}$ decreases quadratically, underscoring the importance of reference-state selection in minimizing experimental overhead.

**3.9 Algorithm Performance Comparison**

Table 4. Performance comparison of quantum eigenenergy computation algorithms, including:
- Maximum evolution time ($T_{max}$)
- Quantum circuit repetitions (total sampling count $N_{tq}$)
- Total evolution time

- Criteria fulfillment:
    (1) Heisenberg-limited precision scaling.
    (2) Use of at most one auxiliary qubit.
    (3) Robustness to sampling errors and circuit noise.
    (4) Additional algorithmic constraints.

| Algorithm | Max Evolution Time | Repetitions | Total Evolution Time | Requirements (1) | (2) | (3) | (4) |
|---|---|---|---|---|---|---|---|
| Standard QPE[48] | $O(\varepsilon^{-1})$ | $O(p^{-1})$ | $O(p^{-1}\varepsilon^{-1})$ | ✓ | ✗ | ✗ | |
| Variant QPE[49–51] | $O(p^{-1}\varepsilon^{-1})$ | $O(p^{-1}\operatorname{polylog}(\varepsilon^{-1}))$ | $O(p^{-2}\varepsilon^{-1})$ | ✓ | ✓ | ? | a |
| Variant QPE[52,53] | $O(\varepsilon^{-1}\log(p^{-1}))$ | $O(p^{-3/2})$ | $O(p^{-3/2}\varepsilon^{-1})$ | ✓ | ✗ | ? | |
| Iterative QPE[54–56] | $O(\varepsilon^{-1})$ | $O(\log(\varepsilon^{-1}))$ | $O(\varepsilon^{-1})$ | ✓ | ✓ | ? | b |
| VQE[57,58] | ? | $O(\varepsilon^{-2})$ | ? | ✗ | ✓ | ✓ | c |
| QITE[59–61] | ? | $O(\varepsilon^{-2})$ | ? | ✗ | ✓ | ✓ | d |
| LCU[62] | $O(p^{-1/2}\varepsilon^{-3/2})$ | $O(p^{-1/2}\varepsilon^{-1/2})$ | $O(p^{-1}\varepsilon^{-2})$ | ✗ | ✗ | ? | |
| Binary-search[63] | $O(p^{-1/2}\varepsilon^{-1})$ | $O(p^{-1/2}\log(\varepsilon^{-1}))$ | $O(p^{-1}\varepsilon^{-1})$ | ✓ | ✗ | ? | |
| QEEA[64] | $O(\varepsilon^{-1}\operatorname{polylog}(p^{-1}))$ | $O(p^{-2}\varepsilon^{-3})$ | $O(p^{-2}\varepsilon^{-4})$ | ✗ | ✓ | ? | |
| Lin et al.[65–67] | $O(\varepsilon^{-1}\operatorname{polylog}(p^{-1}))$ | $O(p^{-2}\operatorname{polylog}(\varepsilon^{-1}))$ | $O(p^{-2}\varepsilon^{-1})$ | ✓ | ✓ | ? | |
| WQTE | $O(\varepsilon^{-1})$ | $O(p^{-2})$ | $O(p^{-2}\varepsilon^{-1})$ | ✓ | ✓ | ✓ | |

a. Precision demands lead to exponential growth in classical computation over measurement data.
b. Requires exact eigenstate ($p = 1$).
c. No guaranteed computational precision.
d. Requires state tomography.

Table 4 presents a comprehensive comparison between our proposed algorithm and existing quantum eigenstate computation methods, including VQE, QITE, LCU, QEEA, and QPE variants. Our algorithm exhibits substantial advantages across five critical

dimensions[68,69]:

(1) **Heisenberg-Limited Precision Scaling**: The proposed algorithm achieves optimal precision scaling of $O(\varepsilon^{-1})$, surpassing the suboptimal scaling of VQE, QITE, LCU, and QEEA. Notably, VQE encounters NP-hard nonlinear non-convex optimization challenges that often impede guaranteed convergence, while QITE necessitates quantum state tomography—a process whose sampling complexity scales exponentially with qubit count. In contrast, our method maintains optimal precision without introducing additional computational bottlenecks.

(2) **Minimal Auxiliary Qubit Requirement**: Whereas standard QPE, LCU, and Binary-search methods require multiple auxiliary qubits (increasing circuit complexity and compromising near-term feasibility), our approach operates efficiently with just a single auxiliary qubit. Although iterative QPE also meets this requirement, it mandates an exact eigenstate as input—a condition often prohibitively resource-intensive to satisfy.

(3) **Enhanced Noise Robustness**: The WQTE method demonstrates remarkable robustness against sampling errors and circuit noise. Theoretical analysis and experimental verification confirm that sampling errors manifest as uniformly distributed white noise in the frequency spectrum, which does not systematically shift the positions of eigenstate peaks, thereby preserving the accuracy of eigenenergy computation. Simultaneously, circuit noise primarily results in overall attenuation of signal amplitudes without distorting the spectral structure; even under practical noisy conditions, moderately increasing the sampling count can effectively restore the peaks suppressed by noise. This dual error tolerance mechanism endows WQTE with the potential for reliable operation on noisy intermediate-scale quantum (NISQ) devices.

(4) **Reduced Maximum Evolution Time**: The maximum evolution time of our method scales comparably to the optimal regime of QPE. In contrast, alternative approaches exhibit dependence not only on precision but also on state overlap, with small overlaps significantly prolonging evolution time.

(5) **Simultaneous Multi-State Computation**: An additional advantage not explicitly

tabulated in Table 4 is our algorithm's capability to compute multiple eigenstates concurrently. This contrasts with methods limited to ground-state calculation (e.g., VQE) or sequential eigenstate estimation.

Among the compared methods, only the approach by Lin et al. approaches the comprehensive performance of our algorithm. However, their method introduces polylogarithmic scaling factors—$\text{polylog}(p^{-1})$ for maximum evolution time and $\text{polylog}(\varepsilon^{-1})$ for repetition count—marginally increasing computational overhead. Furthermore, the noise robustness of their method remains unverified. It is important to clarify that the WQTE algorithm is designed to estimate eigenenergies rather than prepare the corresponding eigenstates. This limitation is inherent to the algorithm's non-variational, spectral-analysis-based framework, which avoids the need for eigenstate preparation but consequently cannot directly output eigenstate wavefunctions.

## 4. Testing WQTE on a Real Quantum Processor

Numerical simulations demonstrate the algorithm's robustness against sampling errors and quantum noise, making it well-suited for current noisy quantum hardware. To validate these results experimentally, we implemented WQTE on the SPINQ Triangulum—a desktop NMR quantum processor based on three $^{19}$F nuclear spins in a $C_2F_3I$ molecule. The device exhibits coherence times of $T_1 \approx 15s$ and $T_2 \approx 500ms$, with a maximum calibrated Rabi rate of 8.3 kHz. It achieves single- and two-qubit gate fidelities of 0.98 and 0.96, respectively, supporting up to ~40 single-qubit gates and ~10 multi-qubit gates per circuit. Unlike superconducting or trapped-ion platforms that require repeated measurements to estimate probability distributions, NMR exploits its ensemble nature (~$10^{23}$ identical molecules) to directly extract expectation values from single-shot macroscopic magnetization readout. While this eliminates the need for multiple sampling (the parameter *M* in Section 3), residual deviations remain due to thermal fluctuations and signal-processing limitations. Given decoherence constraints, the practical circuit depth is limited to several dozen gates.

We tested the system with a minimal 2-site Heisenberg model ($J = 1 \ h = 2$). The corresponding quantum circuit is shown in Fig. 15(a). Rather than employing Trotter decomposition to construct the $e^{-i\hat{H}t}$ operator, we designed the circuit inside the dashed box of Fig. 15(a), which is equivalent to $e^{-i\hat{H}t}$ for the two-qubit Heisenberg model up to a global phase factor $e^{i\cdot 2t}$. To compensate for this phase, we inserted a PhaseShift gate ($PS(\theta) = \begin{pmatrix} 1 & 0 \\ 0 & e^{i\theta} \end{pmatrix}$) on the ancillary qubit.

Using $\Delta = 0.75$ and $N = 800$, Figure 15(b) compares the ideal simulated $q(n)$ signals with the noisy experimental data. While the time-domain traces show noticeable deviations (red vs. black curves), their Fourier spectra (Figure 15(c)) reveal that the errors primarily manifest as uniform white noise, leaving the eigenstate peak positions perfectly intact. This confirms the analysis of Section 3: even under real-device noise, the algorithm preserves spectral accuracy.

It is worth noting that, because only a single circuit sampling was performed per time point in these experiments, the potential mitigation of circuit noise through repeated sampling was not explored. As visible in Fig. 15(c), the spectral peak heights obtained on the real processor are indeed lower than the ideal noise-free case. Fortunately, the circuit complexity was moderate, so the noise impact remained limited and did not suppress the characteristic peaks below the background level of the sampling-induced white noise. Consequently, all eigenstates could still be clearly identified without any loss of accuracy.

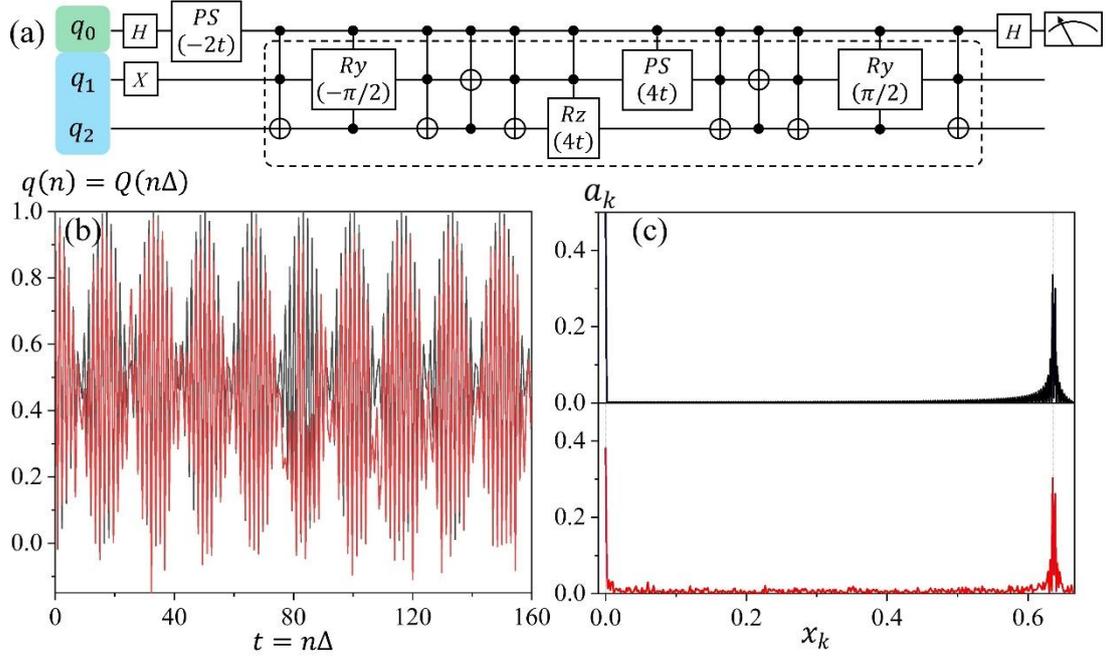

Figure 15. (a) Quantum circuit for the two-qubit Heisenberg model implemented on the NMR quantum processor to test the WQTE algorithm. (b) Time-domain $q(n)$ signals and (c) their corresponding DFT spectra. Black curves: noise-free simulation; red curves: experimental data; blue dashed lines: theoretical eigenfrequencies.

## 5. Conclusion

The WQTE algorithm introduces a non-variational approach to quantum eigen-energy computation, overcoming key limitations of existing methods such as reliance on precise state preparation, high resource demands, and sensitivity to noise. By combining a single-ancilla control scheme with Fourier-based spectral analysis, the algorithm achieves Heisenberg-limited precision while enabling parallel resolution of multiple eigen-energies, significantly reducing circuit complexity. Theoretical and numerical analyses confirm its robustness against sampling errors and quantum noise, with experimental validation on an NMR quantum processor demonstrating practical feasibility in noisy environments. Unlike variational or iterative QPE, WQTE operates efficiently with only a non-zero reference-state overlap, eliminating the need for costly ansatz optimization or exact eigenstate initialization. Its spectral noise resilience ensures reliable eigenvalue extraction even under hardware imperfections, making it particularly suitable for NISQ-era applications. Future improvements could focus on exploring efficient implementations of real-time evolution operators on quantum computers to further reduce the circuit complexity of this algorithm. Overall, WQTE provides a scalable and efficient solution for quantum simulations in chemistry and materials science, bridging the gap between theoretical quantum advantage and near-term experimental implementations.

## Acknowledgments

This work is supported in part by the National Natural Science Foundation of China (Grant Nos. 12504267, 22273069), the Natural Science Foundation of Hubei Province (Grant No. 2025AFB274), and the CPS-Yangtze Delta Region Industrial Innovation Center of Quantum and Information Technology-MindSpore Quantum Open Fund. The source code and experimental data supporting the findings of this study are openly available in the Gitee repository at https://gitee.com/xie-qingxing/quantum-algorithm-wqte under the Apache License 2.0.